\documentclass[11pt,a4paper]{article}
\usepackage{jheppub} 

\usepackage[normalem]{ulem}

\usepackage{pstricks,pst-node,pst-tree}

\usepackage{epic}
\usepackage{mathrsfs}
\usepackage{ae} 
\usepackage[T1]{fontenc}
\usepackage[ansinew]{inputenc}
\usepackage{comment}
\usepackage{ulem}
\usepackage[export]{adjustbox}
\usepackage[english]{babel}
\usepackage{array}
\usepackage{multirow}

\usepackage{float}
\usepackage[font={small}]{caption}
\usepackage{subcaption}

\usepackage{bbold}

\usepackage{tabularx}

\usepackage{cleveref}

\newcommand{\beq}{\begin{equation}}
\newcommand{\eeq}{\end{equation}}
\newcommand{\beqq}{\begin{equation*}}
\newcommand{\eeqq}{\end{equation*}}
\newcommand\beqa{\begin{eqnarray}}
\newcommand\eeqa{\end{eqnarray}}
\newcommand\beqaa{\begin{eqnarray*}}
\newcommand\eeqaa{\end{eqnarray*}}
\newcommand\bea{\begin{array}}
\newcommand\eea{\end{array}}

\def\({\left(}
\def\){\right)}
\def\[{\left[}
\def\]{\right]}

\def\<{\langle}
\def\>{\rangle}

\makeindex

\usepackage{tikz}
\usetikzlibrary{arrows}
\usetikzlibrary{arrows.meta}
\usetikzlibrary{decorations.pathmorphing}
\usetikzlibrary{snakes}
\usetikzlibrary{decorations.pathreplacing,decorations.markings,decorations.pathmorphing}
\usetikzlibrary{decorations.pathreplacing,decorations.markings,decorations.pathmorphing}
\usetikzlibrary{calc}

\title{\boldmath On the space of $U(N)$ scattering amplitudes}

\author[a]{Luc\'ia C\'ordova,}
\affiliation[a]{Department of Theoretical Physics, CERN, 1211 Meyrin, Switzerland}
\emailAdd{lucia.gomez.cordova@cern.ch}
\author[b]{Ricardo Rodrigues}
\affiliation[b]{Centro de F$\acute{\imath}$sica do Porto, Departamento de F$\acute{\imath}$sica e Astronomia,
			Faculdade de Ci$\hat{e}$ncias da Universidade do Porto, Rua do Campo Alegre 687, 4169-007 Porto, Portugal}

\emailAdd{up201805119@edu.fc.up.pt}

\abstract{We investigate the space of massive two-dimensional theories with a global $U(N)$ symmetry and no bound states. Following S-matrix bootstrap principles, we establish rigorous bounds on the space of consistent $2\rightarrow2$ scattering amplitudes. The allowed regions exhibit rich geometric features with integrable models appearing at special points along the boundary. Generic extremal amplitudes display an infinite number of resonances and periodic behavior in energy, similar to previous studies with other group-like symmetries. Within the allowed space, we identify a subregion where the symmetry is enhanced to $O(2N)$, establishing a connection with earlier studies. We also revisit the classification of integrable solutions, identifying one that was previously overlooked in the literature. Finally, we examine the walking behavior of the central charge associated with several of these periodic amplitudes.\\[1.0cm]
\begin{center}
\includegraphics[width=0.45\textwidth]{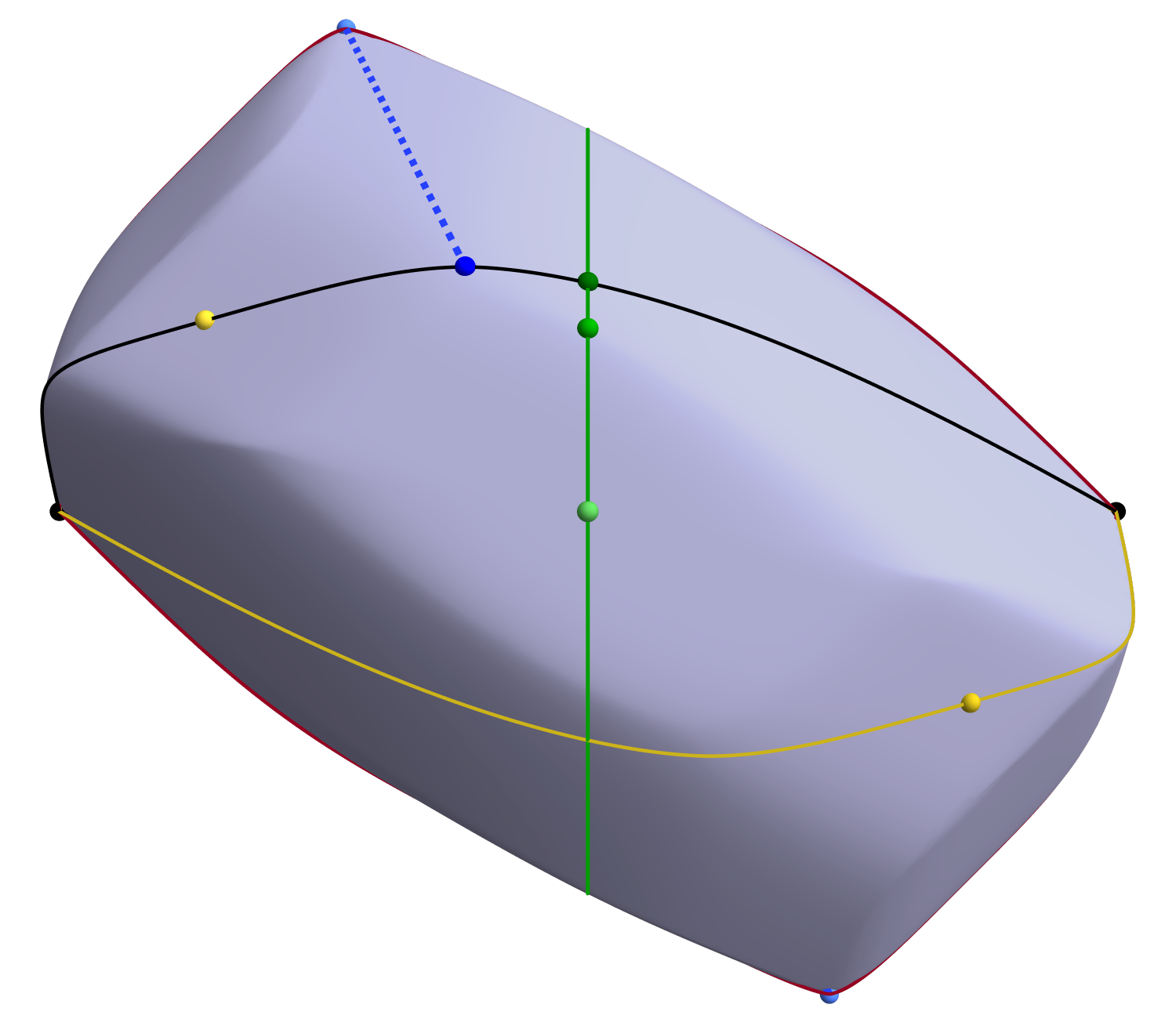}\\[0.3cm]
\textit{U(2) monolith}
\end{center}}

\makeatletter
\gdef\@fpheader{}
\makeatother
\begin{document}
\maketitle

\section{Introduction} \label{sec:Introduction}

The principles of analyticity, crossing symmetry, and unitarity of the S-matrix have proven extremely effective in constraining the space of consistent Quantum Field Theories. In contrast with the original S-matrix program of the 1960s, which aimed at finding a unique consistent S-matrix describing strong interactions, the modern S-matrix bootstrap \cite{Paulos:2016fap,Paulos:2016but} approach has revealed a much richer landscape of possible theories. Particularly simple and instructive examples arise in two-dimensional gapped theories with a global symmetry \cite{He:2018uxa,Cordova:2018uop,Cordova:2019lot,Bercini:2019vme,Homrich:2019cbt,Copetti:2024dcz}, where the bootstrap constraints are already too complex to be solved analytically, making numerical analysis necessary.

An advantage of working in two spacetime dimensions is that one can make contact with integrable models, which offer the rare gift of exact non-perturbative amplitudes for interacting theories. Also, at a more technical level, the simplicity of two-dimensional kinematics allows for an in-depth analysis of extremal amplitudes --i.e. those saturating the bootstrap bounds-- obtained numerically. In particular, these $2\rightarrow2$ amplitudes are parametrized by a small number (as many as irreps in the tensor product of the incoming particles) of functions of a single complex variable $s$, the center of mass energy, whose analytic structure in the complex plane can be thoroughly explored, even at large energies.

In this work, we explore the space of consistent two-dimensional massive theories with a global continuous symmetry $U(N)$. A previous study \cite{Paulos:2018fym} established bounds on cubic couplings in theories with bound states, where the $U(N)$ Gross-Neveu model appears as an extremal amplitude. However, general bootstrap bounds for theories without bound states, as well as a systematic understanding of the different classes of integrable solutions within the space of amplitudes, remained unexplored. We address this gap by deriving rigorous bounds on the possible values of amplitudes at the crossing-symmetric point, using the dual approach formulated in \cite{Cordova:2019lot}.

Our results show integrable physical theories, such as the integrable ``massive" $CP^{N-1}$ model studied in \cite{Basso:2012bw}, the $O(2N)$ non-linear sigma model (NLSM), and free theories, all sitting at vertices of the allowed space. The remaining classes of integrable solutions with periodic behavior in energy, one of which was previously overlooked in the literature, also appear at the boundary of this space. We also find that the rest of extremal amplitudes generically have an infinite number of resonances and exhibit periodic behavior in real rapidity.

The paper is organized as follows: in section~\ref{sec:U(N) S-matrix} we review the necessary group theory concepts to define the amplitudes under consideration, as well as the physical principles used to bootstrap them. We also explain the numerical implementation used to derive bounds. In section~\ref{sec:Numerical results} we present our results, showing various subsections of the space of allowed $U(N)$ amplitudes, along with a full classification of integrable amplitudes that live in this space. We also discuss various features of the extremal amplitudes. Finally in section~\ref{sec:Discussion} we examine the walking behavior of the UV central charge for periodic amplitudes, hinting to the presence of complex conformal theories.

Several appendices complement this work: in appendix~\ref{app: YB solutions} we present a thorough review of the classification of integrable amplitudes, in appendix~\ref{app:form factor expressions} we give more details on the integrable form factors used to compute the central charges, and in appendix~\ref{app:codes} we provide examples of the dual and primal bootstrap codes used to derive numerical bounds.

\section{Preliminary notions} \label{sec:U(N) S-matrix}
In this section we will address the structure of the S-matrix for the scattering of $U(N)$ charged particles of mass $m$ and the conditions which constrain it. In particular, we focus on two body scattering of particles transforming in the fundamental representation.

Given that the symmetry group $U(N)$ is complex, this implies that we will have particles, transforming under an irreducible representation $\mathcal{R}$, as well as antiparticles transforming under the corresponding complex conjugate irrep $\overline{\mathcal{R}}$. As such, there are two physically distinct scenarios: particle-particle and particle-antiparticle scattering. For these two cases the two-dimensional S-matrix can be written as:
\begin{align}
    & \langle P_{\bar{\delta}} (p_4) P_{\bar{\gamma}} (p_3) | \, \mathcal{S} \, | P_{\alpha} (p_1) P_{\beta} (p_2) \rangle = S_{\alpha \, \beta}^{\overline{\delta} \, \overline{\gamma}}(\theta) \left[ \delta_{k_1,k_3}  \delta_{k_2,k_4} \pm  \delta_{k_1,k_4}  \delta_{k_2,k_3} \right] \, , \label{eq:PP S-matrix}\\
    & \langle A_{\delta} (p_4) P_{\bar{\gamma}} (p_3) | \, \mathcal{S} \, | P_{\alpha} (p_1) A_{\bar{\beta}} (p_2) \rangle = F_{\alpha \, \overline{\beta}}^{\delta \, \bar{\gamma}}(\theta) \, \delta_{k_1,k_3}  \delta_{k_2,k_4} + B_{\alpha \,  \overline{\beta}}^{\overline{\gamma} \,  \delta}(\theta) \, \delta_{k_1,k_4}  \delta_{k_2,k_3} \, , \label{eq:PA S-matrix}
\end{align}
where $P_{\alpha}$, $P_{\bar{\alpha}}$ denote respectively ingoing and outgoing particle states whereas $A_{\bar{\beta}}$ and $A_{\beta}$ designate ingoing and outgoing antiparticle states, and $p_i^{\mu} = (E_i,k_i)$. Furthermore, the rapidity variable $\theta$ is related with the usual Mandelstam invariant $s = (p_1 + p_2)^2$ by $s = 4 \, m^2 \, \cosh^2\left( \theta /2\right)$. In the first equation the $\pm$ sign accounts for the bosonic $(+)$ or fermionic $(-)$ nature of the particles. Moreover, $F_{\alpha \, \overline{\beta}}^{\delta \, \overline{\gamma}}(\theta)$ denotes the forward amplitude whereas $B_{\alpha \, \overline{\beta}}^{\overline{\gamma} \, \delta}(\theta)$ is the backward amplitude. All three functions in the above equations are related to each other by crossing symmetry
\begin{align}
    & S_{\alpha \, \beta}^{\overline{\delta} \, \overline{\gamma}}(i \pi -  \theta) = F_{\alpha  \, \overline{\delta}}^{\beta  \, \overline{\gamma}}(\theta) \quad , \quad B_{\alpha \, \overline{\beta}}^{\overline{\gamma} \, \delta}(i \pi - \theta) = B_{\delta \, \overline{\beta}}^{\overline{\gamma} \, \alpha}(\theta) \, ,
    \label{eq:crossing conditions for S,F,B}
\end{align}
which translates the fact that by swapping ingoing and outgoing particles in one scattering process we can get a different one.

As found in \cite{Berg:1977dp} we can write the functions $S$, $F$ and $B$ in terms of $U(N)$ invariant amplitudes
\begin{align}
    & S_{\, \alpha \, \beta}^{\, \overline{\delta} \, \overline{\gamma}} (\theta) =  \vcenter{\hbox{\includegraphics[scale=0.92]{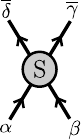}}} = u_1 (\theta) \, \vcenter{\hbox{\includegraphics[scale=0.92]{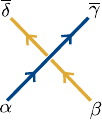}}} + u_2 (\theta) \, \vcenter{\hbox{\includegraphics[scale=0.92]{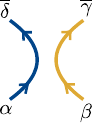}}} \, , \label{eq:S matrix def} \\
    & F_{\, \alpha \, \overline{\beta}}^{\, \delta \, \overline{\gamma}} (\theta) = \vcenter{\hbox{\includegraphics[scale=0.92]{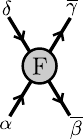}}} = t_1 (\theta) \, \vcenter{\hbox{\includegraphics[scale=0.92]{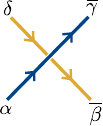}}} + t_2 (\theta) \, \vcenter{\hbox{\includegraphics[scale=0.92]{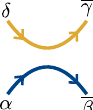}}} \, , \label{eq:F matrix def}\\
    & B_{\, \alpha \, \overline{\beta}}^{\, \overline{\gamma} \, \delta}    (\theta) = \vcenter{\hbox{\includegraphics[scale=0.92]{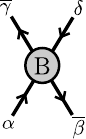}}} = r_1 (\theta) \, \vcenter{\hbox{\includegraphics[scale=0.92]{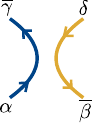}}} + r_2 (\theta) \, \vcenter{\hbox{\includegraphics[scale=0.92]{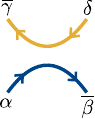}}} \, , \label{eq:B matrix def}
\end{align}
where each diagram represents the product of two Kronecker deltas.

In this basis the crossing relations are straightforward. Indeed, by looking at the pictorial representations it is easy to see that some of diagrams are related by a rotation of $90^{\circ}$. Since this rotation amounts to sending $\theta \rightarrow i \pi - \theta$, it easily follows that crossing translates into
\begin{align}
    u_1(\theta) = t_1(i \pi - \theta) \, , \qquad u_2(\theta) = t_2(i \pi - \theta) \, , \qquad r_1(\theta) = r_2(i \pi - \theta) \, .
    \label{eq:crossing conditions for u,t,r}
\end{align}

Alternatively the S-matrix can be decomposed in terms of the representations of $U(N)$ that appear in the product of the fundamental representation $N$ with itself or with its complex conjugate $\overline{N}$:
\begin{align}
    &N \otimes N = \frac{N(N+1)}{2} \oplus \frac{N(N-1)}{2}  \quad , \quad N \otimes \overline{N} = 1 \oplus (N^2 - 1) \, ,
    \label{eq:irrep decomposition for U(N) lixo}
\end{align}
where on the left equation the representations correspond to the symmetric and antisymmetric representations respectively, whereas on the right one we encounter the singlet and the adjoint representations. In this basis we have
\begin{align}
    &S = S_{\text{sym}}(\theta) \, \mathbb{T}_{\text{sym}} + S_{\text{anti}}(\theta) \, \mathbb{T}_{\text{anti}} \, , \label{eq:S in U_N reps basis} \\
    &F^{+} \equiv F + B = S_{\text{sing}}^{+} (\theta) \, \mathbb{T}_{\text{sing}} +  S_{\text{adj}}^{+} (\theta) \, \mathbb{T}_{\text{adj}} \, , \label{eq:F plus in U_N reps basis}\\
    &F^{-} \equiv F - B = S_{\text{sing}}^{-} (\theta) \, \mathbb{T}_{\text{sing}} +  S_{\text{adj}}^{-} (\theta) \, \mathbb{T}_{\text{adj}} \, , \label{eq:F minus in U_N reps basis}
\end{align}
where we have defined $F^{\pm}$ as in \cite{Paulos:2018fym}, with $\pm$ denoting the parity, and with the representations' structures given by
\begin{align}
        & (\mathbb{T}_{\text{sym}})_{\alpha  \beta}^{\overline{\delta}  \overline{\gamma}} = \frac{1}{2}\left( \delta_{\alpha}^{\overline{\gamma}} \,  \delta_{\beta}^{\overline{\delta}} + \delta_{\alpha}^{\overline{\delta}} \,  \delta_{\beta}^{\overline{\gamma}} \right) \quad , \,  &(\mathbb{T}_{\text{anti}})_{\alpha  \beta}^{\overline{\delta}  \overline{\gamma}} = \frac{1}{2}\left( \delta_{\alpha}^{\overline{\gamma}} \,  \delta_{\beta}^{\overline{\delta}} - \delta_{\alpha}^{\overline{\delta}} \,  \delta_{\beta}^{\overline{\gamma}} \right) \, ,\\
    & (\mathbb{T}_{\text{sing}})_{\alpha  \overline{\beta}}^{\delta  \overline{\gamma}} = \frac{1}{N} \, \delta_{\alpha \overline{\beta}} \, \delta^{\delta \overline{\gamma}} \quad , \,  &(\mathbb{T}_{\text{adj}})_{\alpha \, \overline{\beta}}^{\delta \, \overline{\gamma}} =  \delta_{\alpha}^{\overline{\gamma}} \,  \delta_{\overline{\beta}}^{\delta} - \frac{1}{N} \, \delta_{\alpha \overline{\beta}} \, \delta^{\delta \overline{\gamma}} \, .
\end{align}

The two basis we introduced are related in the following way:
\begin{equation}
   \begin{aligned}
       S_{\text{sym}} &= u_1 + u_2 \, , &S_{\text{anti}} = u_1 - u_2 \, , \\
       S_{\text{sing} \, \pm} &= t_1 \pm r_1 + N \, (t_2 \pm r_2) \, ,\qquad  &S_{\text{adj} \, \pm} = t_1 \pm r_1 \, .
    \end{aligned}
    \label{eq:change of basis}
\end{equation}

Using this map it is possible to reexpress the crossing conditions in (\ref{eq:crossing conditions for u,t,r}) in terms of the functions $S_{a}$ ($a \in$ sing $\pm$, adj $\pm$, sym, anti) as
\begin{align}
    S_a(i \pi - \theta) = \mathcal{C}_{a \, b} \, S_{b}(\theta) \, .
    \label{eq:crossing relation for S}
\end{align}
with
\begin{align}
    & \mathbf{S}(\theta) \equiv \begin{pmatrix} S_{\text{sing} +} (\theta) \\ S_{\text{adj} +} (\theta) \\ S_{\text{sing} -} (\theta) \\ S_{\text{adj} -} (\theta) \\ S_{\text{sym}} (\theta) \\ S_{\text{anti}}(\theta)\end{pmatrix} \, ,
    \label{eq:S matrix components vector form}
\end{align}
and where the crossing matrix $\mathcal{C}_{a\, b}$ is given by
\begin{align}
    \mathcal{C}_{a \, b} \equiv \left(\begin{array}{cccccc}
   \frac{1}{2 N}&\frac{N^2 - 1}{2N}&-\frac{1}{2 N}&\frac{1- N^2}{2 N}&\frac{1+N}{2}&\frac{1-N}{2}
   \\[3pt] \frac{1}{2 N}& - \frac{1}{2 N}&-\frac{1}{2 N}&\frac{1}{2 N}&\frac{1}{2}&\frac{1}{2}
   \\[3pt] - \frac{1}{2 N}&\frac{1 - N^2}{2N}&\frac{1}{2 N}&\frac{N^2 - 1}{2 N}&\frac{1+N}{2}&\frac{1-N}{2}
   \\[3pt] - \frac{1}{2 N}& \frac{1}{2 N}&\frac{1}{2 N}& - \frac{1}{2 N}&\frac{1}{2}&\frac{1}{2} 
   \\[3pt] \frac{1}{2N}& \frac{N-1}{2 N}& \frac{1}{2 N}& \frac{N - 1}{2 N}& 0& 0
   \\[3pt] - \frac{1}{2 N}& \frac{1 + N}{2 N}& - \frac{1}{2 N}& \frac{1 + N}{2 N}& 0 & 0 \end{array}\right) \, ,
   \label{eq:full U(N) crossing matrix}
\end{align}
such that it satisfies $\mathcal{C}^2 = \mathbb{1}$. 

In the basis of (\ref{eq:S in U_N reps basis}), (\ref{eq:F plus in U_N reps basis}) and (\ref{eq:F minus in U_N reps basis}), the unitarity conditions for the S-matrix translate into
\begin{align}
    &|S_{a} (\theta)|^2 \le 1 \qquad \text{($a$ = sing $\pm$, adj $\pm$, sym, anti)} \, ,
    \label{eq:unitarity condition in representation components}
\end{align}
for $\theta \ge 0$ in the real axis, i.e. $s\ge4m^2$. These are somewhat simpler than if expressed in the $u_i, \, t_i, \, r_i$ $(i=1,2)$ basis, which can be easily obtained from the above inequalities by making use the map (\ref{eq:change of basis}).


\subsection{Numerical implementation}\label{subsec:numerics}
We shall now say a few words about how the space of the allowed amplitudes is mapped in practice. The reader familiar with S-matrix bootstrap can safely skip this section and proceed to the next one.

Regardless of the approach one follows, when bootstrapping the S-matrix we must always bear in mind and impose the three consistency conditions previously mentioned:

\begin{figure}[t]
    \centering
    \includegraphics[scale = 1.3]{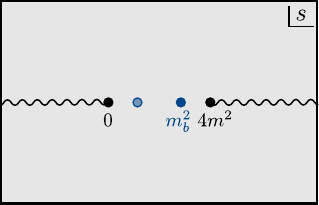}
    \caption{Analytic structure of the S-matrix function $S(s)$ in the complex plane $s\in \mathbb{C}$. Poles are connected with the existence of bound states whereas branch cuts represent multi-particle thresholds. }
    \label{fig:analytic structure in s complex plane}
\end{figure}

\begin{itemize}
    \item We assume that the S-matrix $\mathbf{S}(s)$ is an analytic function in the $s$ complex plane apart from possible poles representing bound states and branch cuts for intermediate multi-particle states, as depicted in figure \ref{fig:analytic structure in s complex plane}. In our case we assume no bound states in the spectrum such that we should only have branch cuts for $s \ge 4 m^2$ and $s \le 0$, where the latter concerns the $t$-channel;
    \item Moreover, we also impose that the amplitude is crossing symmetric, as the s-and t-channel processes are boundary values of the same analytic function. For the simplest case of $2 \rightarrow 2$ scattering of identical particles this translates into $S(s) = S(t=4 m^2 - s)$. In our case where we have a certain global symmetry and consequently a function for each representation of the group, this condition is slightly more convoluted (\ref{eq:crossing relation for S});
    \item Lastly, the full S-matrix must also satisfy the unitarity condition, namely that the S-matrix operator is an unitary operator $\mathcal{S} \, \mathcal{S}^\dagger = \mathbb{1}$. From a quantum mechanics perspective this is simply the statement of probability conservation. For the case mentioned before, specifically the scattering of identical particles with no global symmetry, the unitarity condition translates to: $|S(s)|^2 \le 1$ for $s \ge 4m^2$. Seeing as in here we have several functions for each representation, each of these must satisfy a similar unitarity condition as expressed in (\ref{eq:unitarity condition in representation components});
\end{itemize}
Based in these conditions, the numerical bootstrap can be carried out in two different ways.

In the primal approach, put forward in \cite{Paulos:2016but}, we start by constructing an ansatz for the S-matrix that automatically satisfies both crossing and analyticity. One such possible choice is
\begin{equation}
    \mathbf{S}(s) = \mathbf{S}_0 + \sum_{k=1}^{\infty} \left[ \mathbf{S}_k \, \rho^k_{2m^2}(s) + \mathcal{C} \cdot \mathbf{S}_k \, \rho^k_{2m^2}(4m^2 - s) \right] \, ,
    \label{eq:rho ansatz for the S-matrix}
\end{equation}
where
\begin{align}
    & \mathbf{S}_n = \begin{pmatrix} \text{sing}^{+}_n \\ \text{adj}^{+}_n \\ \text{sing}^{-}_n \\ \text{adj}^{-}_n \\ \text{sym}_n \\ \text{anti}_n  \end{pmatrix} \, , \quad n \in \mathbb{N}_0
    \label{eq:S matrix Taylor expansion coefficients vector form}
\end{align}
is a vector of coefficients for this Taylor-like expansion of each one of the components of the S-matrix, and the variable
\begin{equation}
    \rho_{s_0} (s) = \frac{\sqrt{4 m^2 - s_0} - \sqrt{4 m^2 - s}}{\sqrt{4 m^2 - s_0} + \sqrt{4 m^2 - s}} \, ,
    \label{eq:rho variable definition}
\end{equation}
maps the right half plane of $s \in \mathbb{C}$ into the unit disk, where in particular $s_0$ gets mapped to its origin, see for example \cite{Hogervorst:2013sma, Paulos:2017fhb}. Crucially, this variable takes care of analyticity as the $s \ge 4m^2$ branch cut is mapped to the boundary of the disk. As for crossing, it is easy to see from (\ref{eq:rho ansatz for the S-matrix}), which was written purposefully in a crossing symmetric way, that it satisfies (\ref{eq:crossing relation for S}). After we construct the ansatz for the S-matrix we choose a given functional $\mathcal{F}[\mathbf{S}(s)]$ which we want to maximize, while simultaneously imposing the unitarity condition. This fixes the coefficients in (\ref{eq:rho ansatz for the S-matrix}) and provides (an approximation) of a consistent scattering amplitude. In practice, unitarity is imposed by evaluating our ansatz at several points $s_i$ ($i=1 , \dots, N_{\text{grid}}$) in the physical region $s \ge 4 m^2$, and demanding that in each of these $|S_a(s_i)|^2 \le 1$ $\text{($a$ = sing $\pm$, adj $\pm$, sym, anti)}$ is satisfied. As for the functionals, some of the usual choices for these correspond to: $\mathcal{F}[\mathbf{S}(s)] = \mathbf{S}(s_*)$ for $0 < s_* < 4m^2$, $\mathcal{F}[\mathbf{S}(s)] = \text{Res}_{s=m_b^2} \, \mathbf{S}(s)$ where $m_b$ is some bound state mass, between others. By optimizing these functionals we bound the space of allowed amplitudes as desired. The benefit of the primal approach is that by the end of the process we end up with an explicit expression of the S-matrix, which in spite of being an approximation, is good enough to allow us to study some of its characteristics. However, there is also a downside to this approach, which is that the bounds are not rigorous and their values depend on the number of parameters we include in the ansatz.

On the other hand we have the dual approach first introduced in \cite{Cordova:2019lot} and later extended to different settings in \cite{Correia:2022dyp, Cordova:2023wjp, Guerrieri:2020kcs, Kruczenski:2020ujw, He:2021eqn, Guerrieri:2021tak, Guerrieri:2024ckc}, which not only places rigorous bounds 
as it is also faster than the primal one. In this method we first optimize a dual functional $\mathcal F_d(\mathbf K(s))$ in terms of dual functions $\mathbf{K}(s)$, which serve as Lagrange multipliers for the S-matrix constraints. To give a concrete example, if we want to put bounds on $S(2m^2)$ the dual function has the following form 
\begin{equation}
   \begin{aligned}
    &\mathbf{K}(s) = \frac{1}{ (s-2m^2) \, \sqrt{s \, (s-4m^2)}} \biggl\{ \sum_{i=1}^{3} a_i \, \mathbf{v_i} \\
    &+ \sum_{j=1}^{\infty} \left[ \mathbf{K}_j \, \rho^j_{2m^2}(s) + \mathcal{C}^T \cdot \mathbf{K}_j \, \rho^j_{2m^2}(4m^2 - s) \right] \biggl\} \, ,
    \end{aligned}
    \label{eq:rho ansatz for K}
\end{equation}
where the coefficients $a_i$ are to be fixed by the optimization procedure, $\mathbf{v_i}$ are eigenvectors of $\mathcal{C}^T$ of eigenvalue $1$ and $ \mathbf{K_j}$ are vectors of coefficients similar to (\ref{eq:S matrix Taylor expansion coefficients vector form}). The initial prefactor is crucial in order to guarantee that we are exploring the space of amplitudes at the crossing symmetric point $s=2m^2$, and at the same time takes care of the desired decay rate as $s\rightarrow\infty$. The function $\mathbf{K}(s)$ also satisfies an anti-crossing condition with respect to the transverse crossing matrix $\mathcal{C}^T$, i.e. $\mathbf{K}(4m^2 - s) = - \mathcal{C}^T \cdot \mathbf{K}(s)$. This can be easily be seen to hold with the above ansatz. The dual functional $\mathcal{F}_d [\mathbf{K}(s)]$ is given by
\begin{equation}
    \mathcal{F}_d [\mathbf{K}(s)] = \frac{2}{\pi} \int_{4m^2}^{\infty} ds \sum_{a} |K_a (s)| \, ,
    \label{eq:dual functional definition}
\end{equation}
where the sum is over the components of $\mathbf{K}(s)$ and where the integration can be numerically achieved with the Chebyshev method, for example. It is also important that each $K_a(s)$ decays at least like $s^{-3/2}$ in order for the integral to converge. Finally, we minimize this functional which by construction satisfies
\begin{equation}
    \max_{\{\mathbf{S}\}} \, \mathcal{F} [\mathbf{S}(s)] \le \min_{\{\mathbf{K}\}} \mathcal{F}_d [\mathbf{K}(s)] \, ,
    \label{eq:relation functional and dual functional}
\end{equation}
implying that we are always approaching the optimal bound from above. One of the reasons why this method is faster when compared with the primal one is that we do not have to impose the unitarity constraints while minimizing the dual functional.

In this work we we used both methods to insure optimality of our bounds. The \texttt{Mathematica} codes for the two approaches can be found in appendix \ref{app:codes}. Finally, we mention that for the minimization / maximization of functionals, we have used the optimization solver library \texttt{MOSEK} \cite{mosek}, which significantly reduces the required time compared to the built-in methods of \texttt{Mathematica}.

\section{The space of allowed amplitudes} \label{sec:Numerical results}

\begin{figure}[ht]
\begin{subfigure}[][][c]
{0.775\textwidth}
\includegraphics[width=.765\textwidth,right,valign=c]{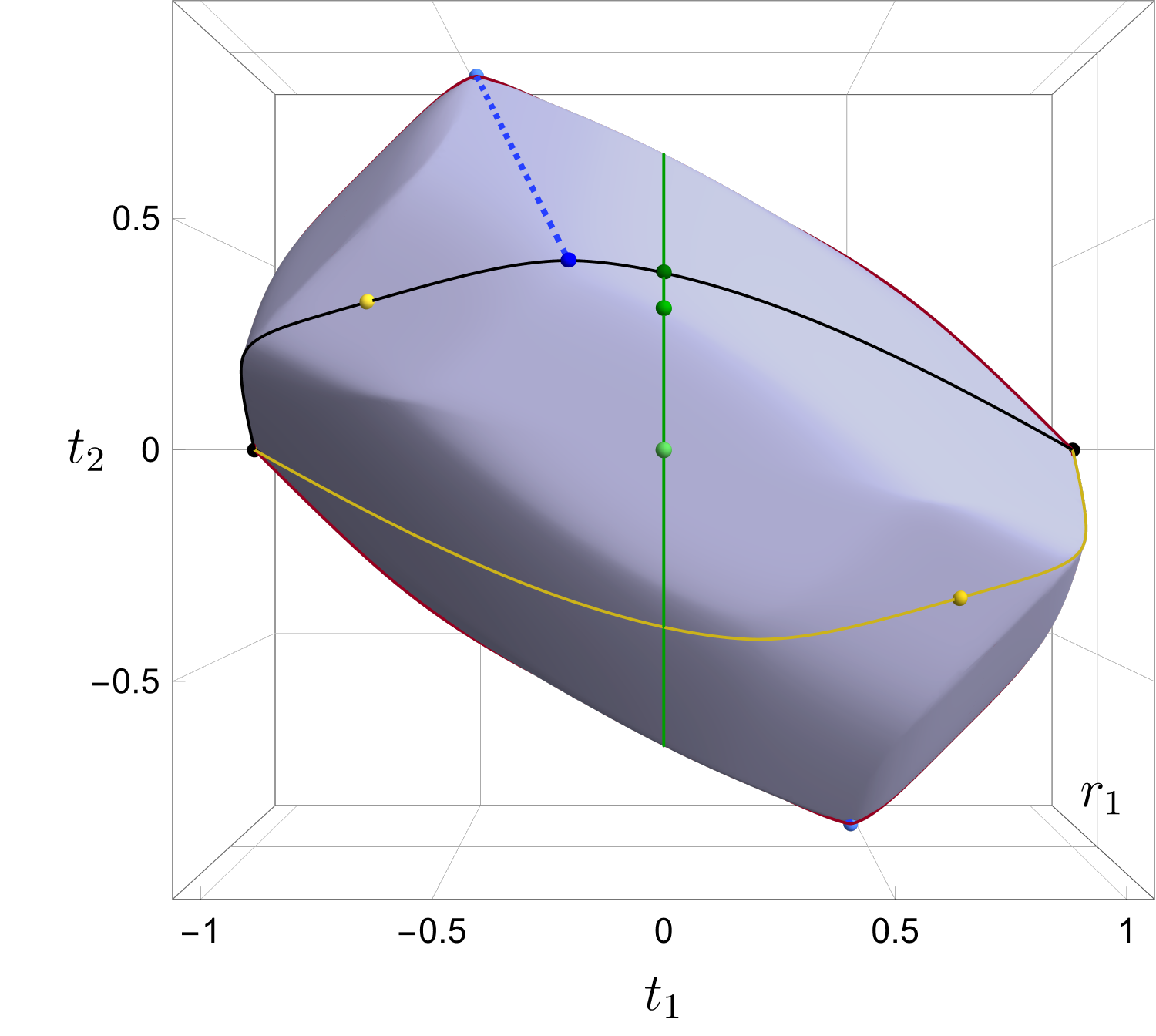}
\end{subfigure}
\raisebox{15pt}{
\begin{subfigure}[][][c]
{0.18\textwidth}
\includegraphics[width=.7\textwidth,left,valign=top]{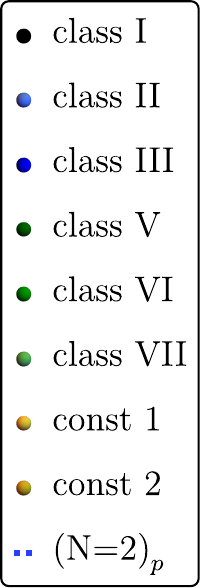}
\end{subfigure}
}
\hfill\vspace{.2cm}
\centering
\begin{subfigure}[][][c]
{0.99\textwidth}\centering
\includegraphics[width=.58\textwidth]{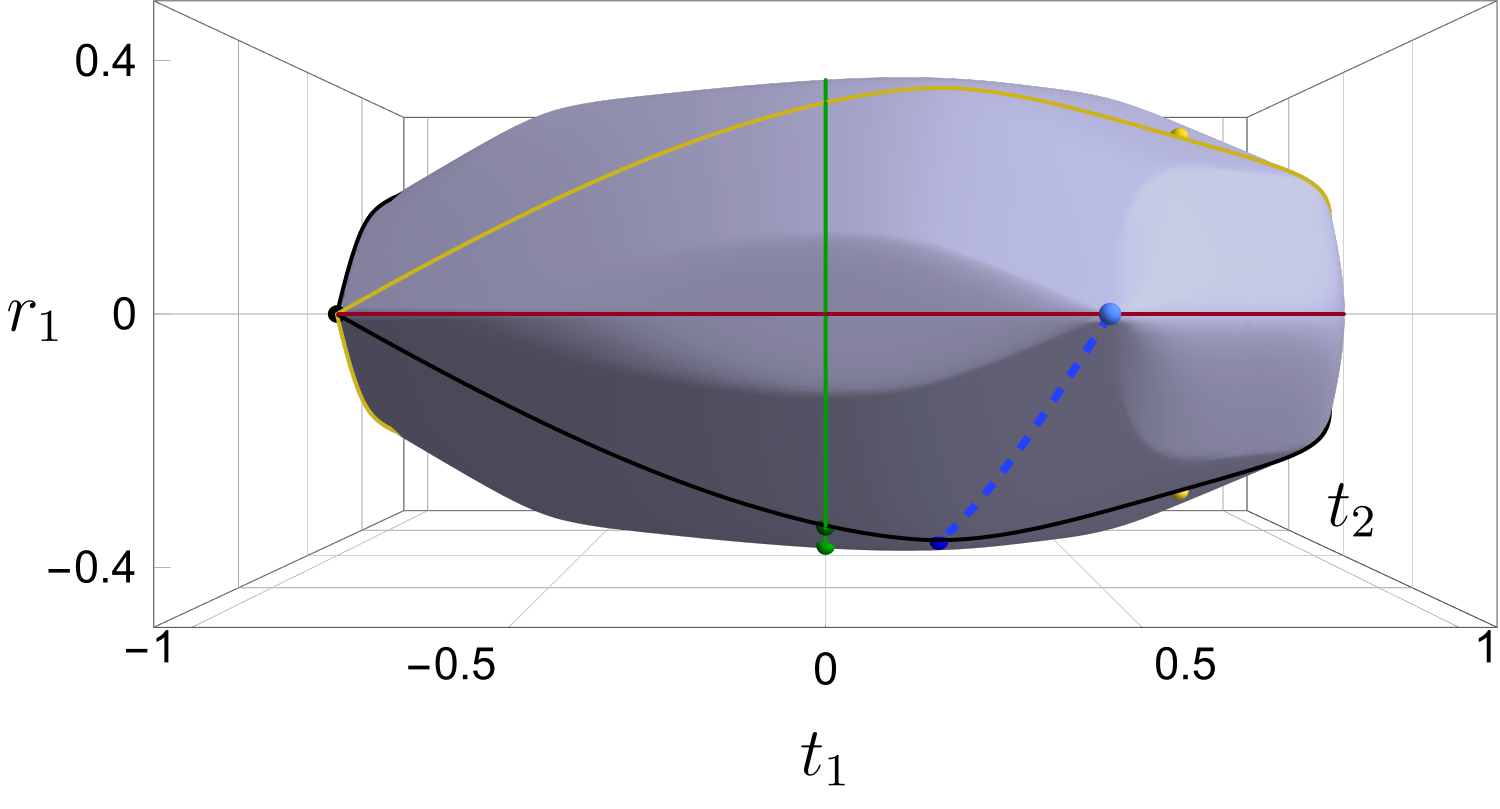}
\end{subfigure}\hfill\vspace{0.2cm}
\begin{subfigure}[][][c]{0.99\textwidth}\centering
\includegraphics[width=.565\textwidth]{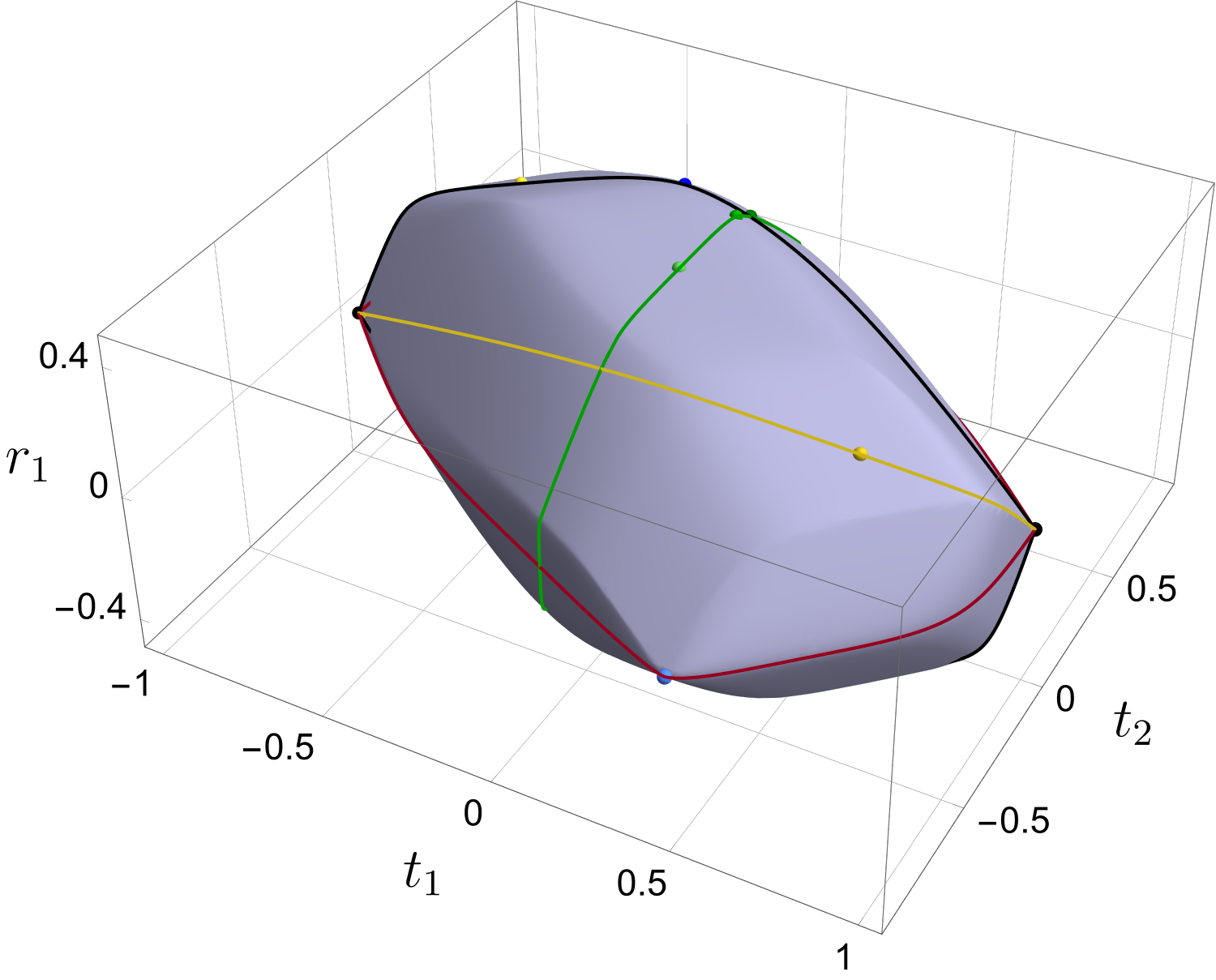}
\end{subfigure}
\caption{$U(2)$ monolith giving the space of allowed amplitudes parametrized by $\{t_1,t_2,r_1\}$ in (\ref{eq:S matrix def}-\ref{eq:B matrix def}) at the crossing symmetric point $\theta=i\pi/2$. Integrable (table~\ref{table:integrable}) and constant solutions \eqref{eq: def const} situated at the boundary are highlighted.}
\label{fig:N2monos}
\end{figure}

\begin{figure}[t]
\begin{subfigure}
{0.5\textwidth}
\includegraphics[width=.97\textwidth,left]{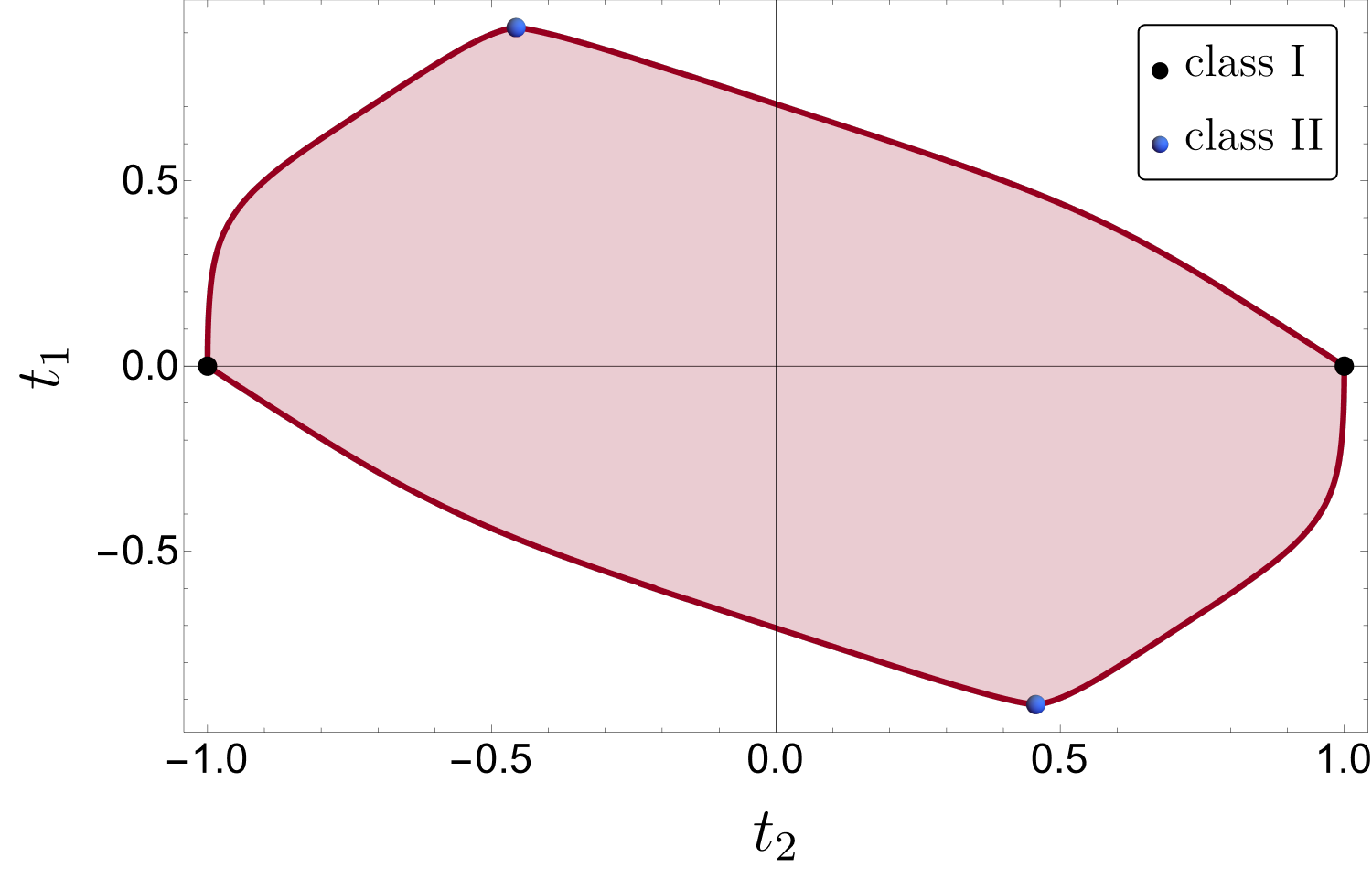}
\caption{$r_1=0$}
\end{subfigure}
\begin{subfigure}
{0.5\textwidth}
\includegraphics[width=.97\textwidth,right]{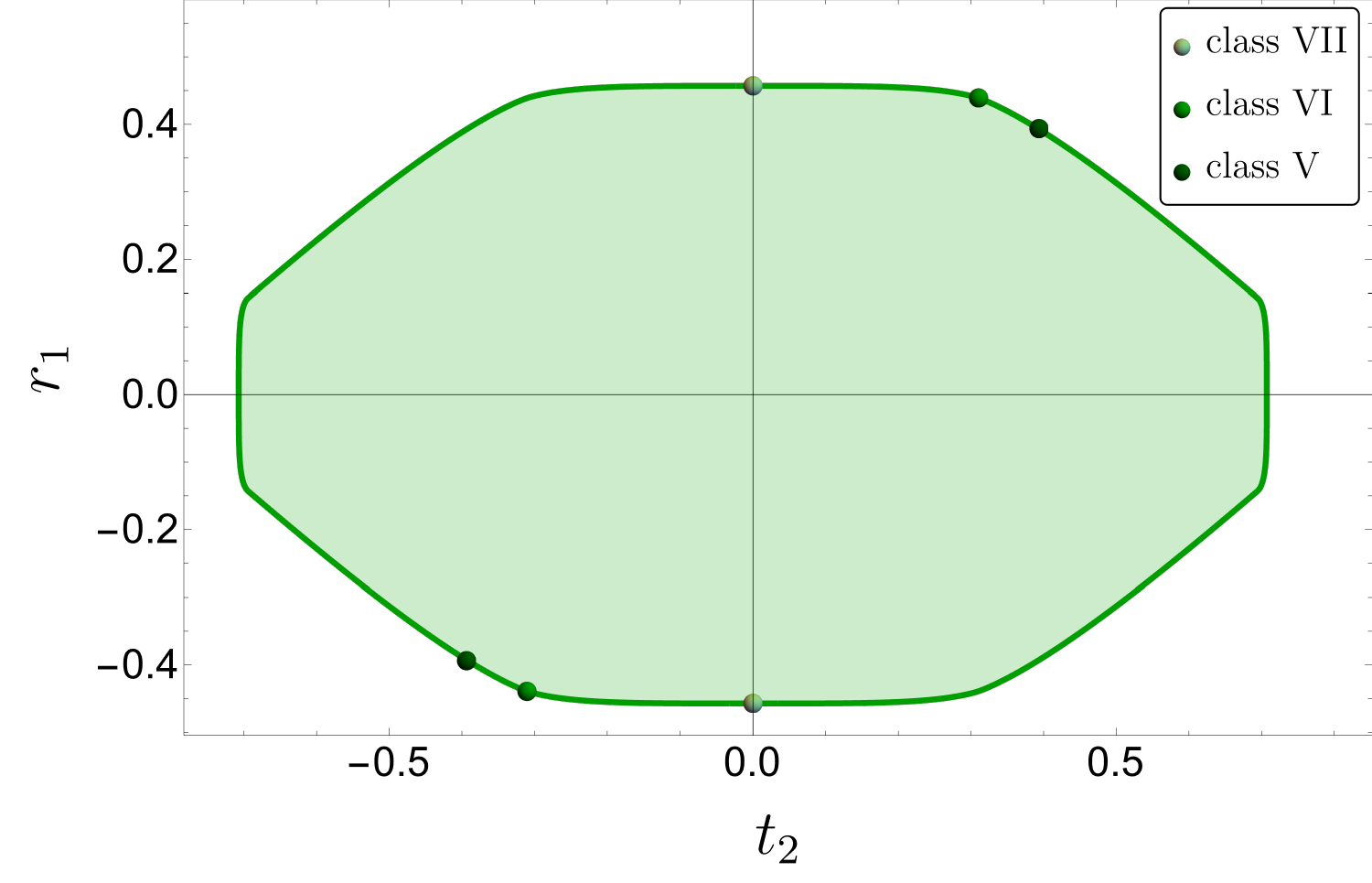}
\caption{$t_1=0$}
\end{subfigure}\vspace{.5cm}
\begin{subfigure}
{0.5\textwidth}
\includegraphics[width=.97\textwidth,left]{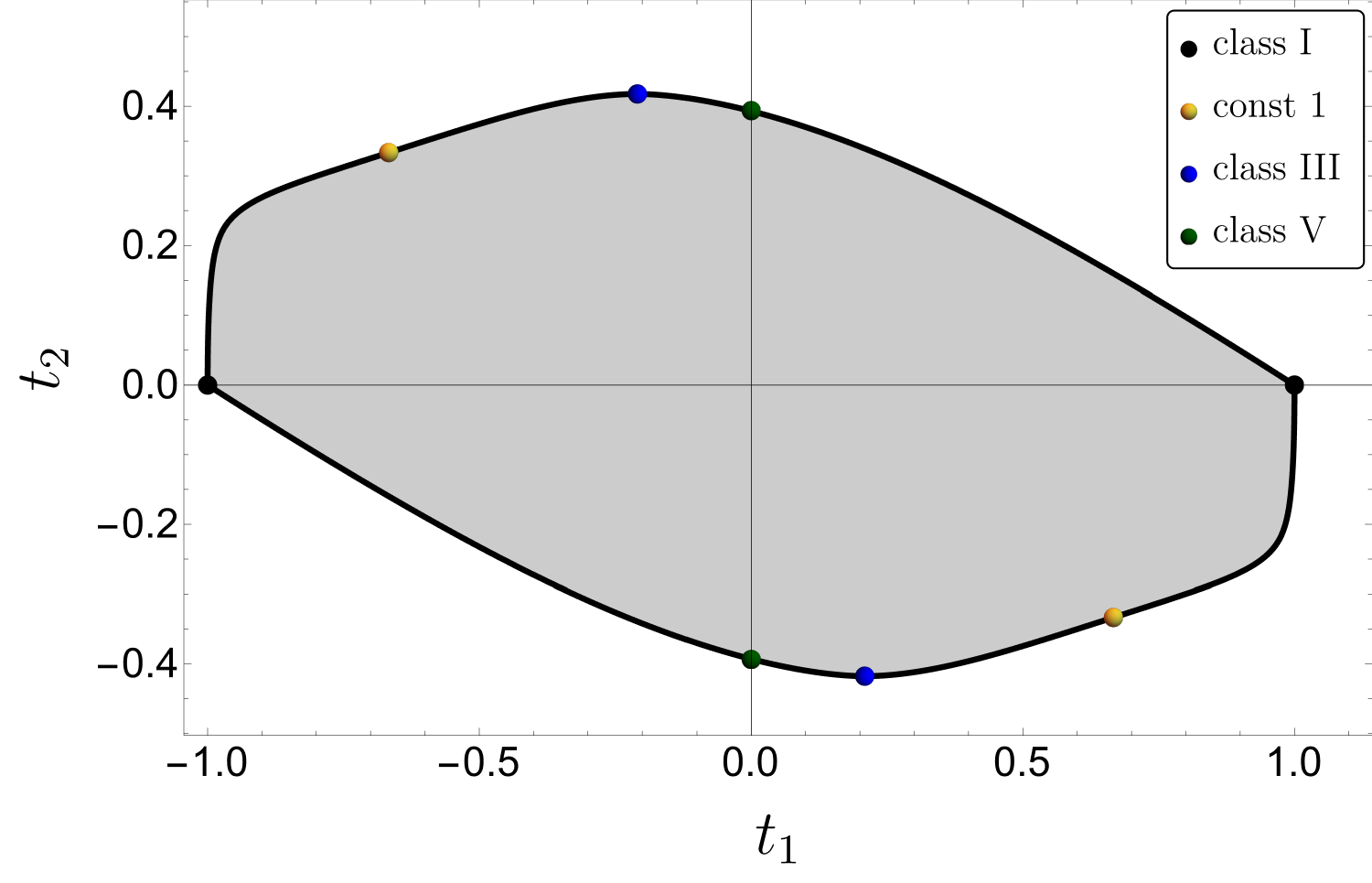}
\caption{$t_2=r_2$}
\end{subfigure}
\begin{subfigure}
{0.5\textwidth}\centering
\includegraphics[width=.97\textwidth,right]{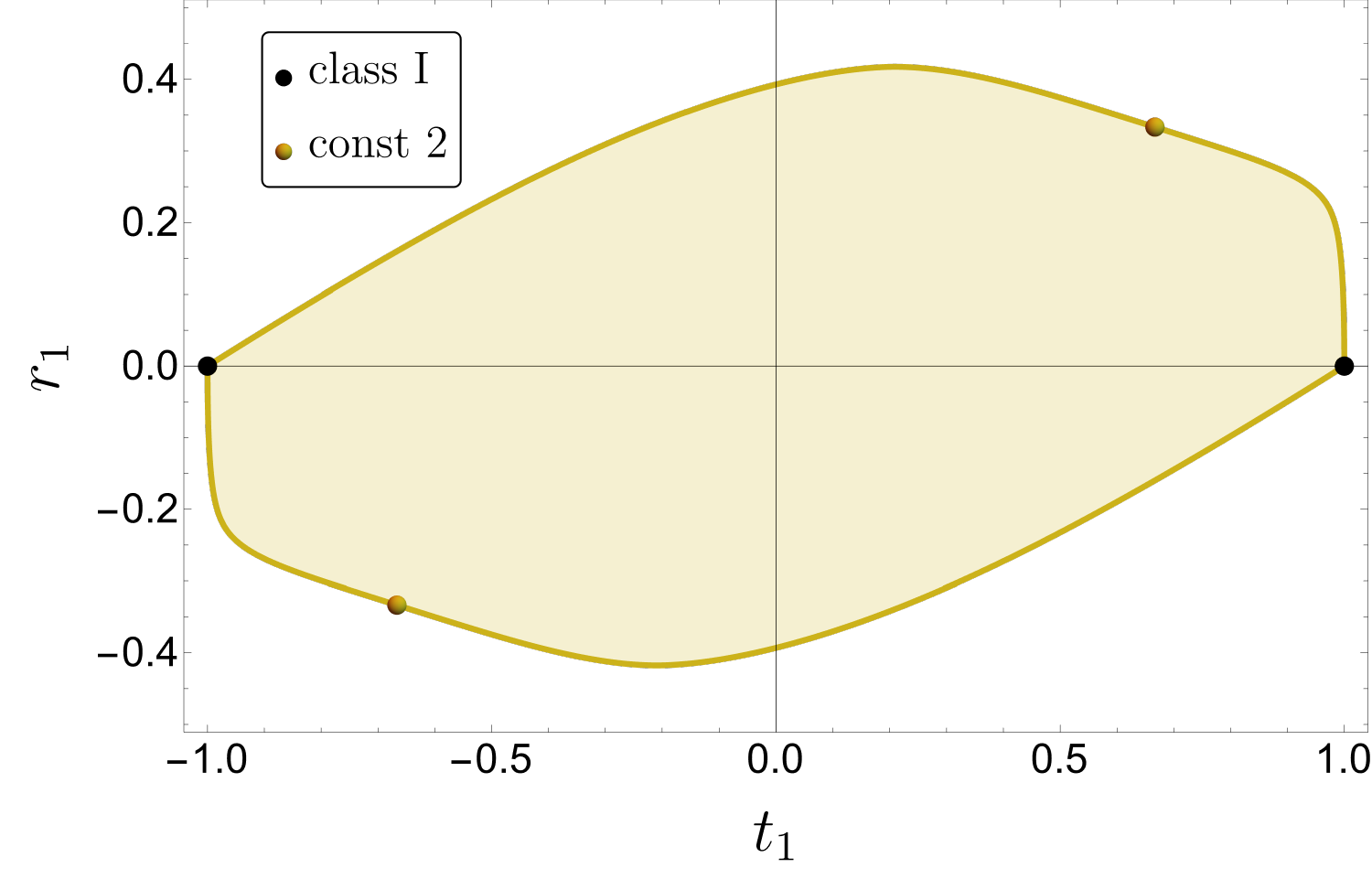}
\caption{$t_2=-r_2$}
\end{subfigure}
\caption{Sections of the $U(2)$ monolith (see solid lines in figure~\ref{fig:N2monos}), obtained with the dual code in appendix \ref{app:codes} using \textit{nmax} = $30$, \textit{nintpts} = $90$, \textit{precision} = $100$ and $\sim1000$ angles.}
\label{fig:N2sections}
\end{figure}

\begin{figure}[th!]
\begin{subfigure}[][][c]
{0.86\textwidth}
\includegraphics[width=.9\textwidth,right,valign=c]{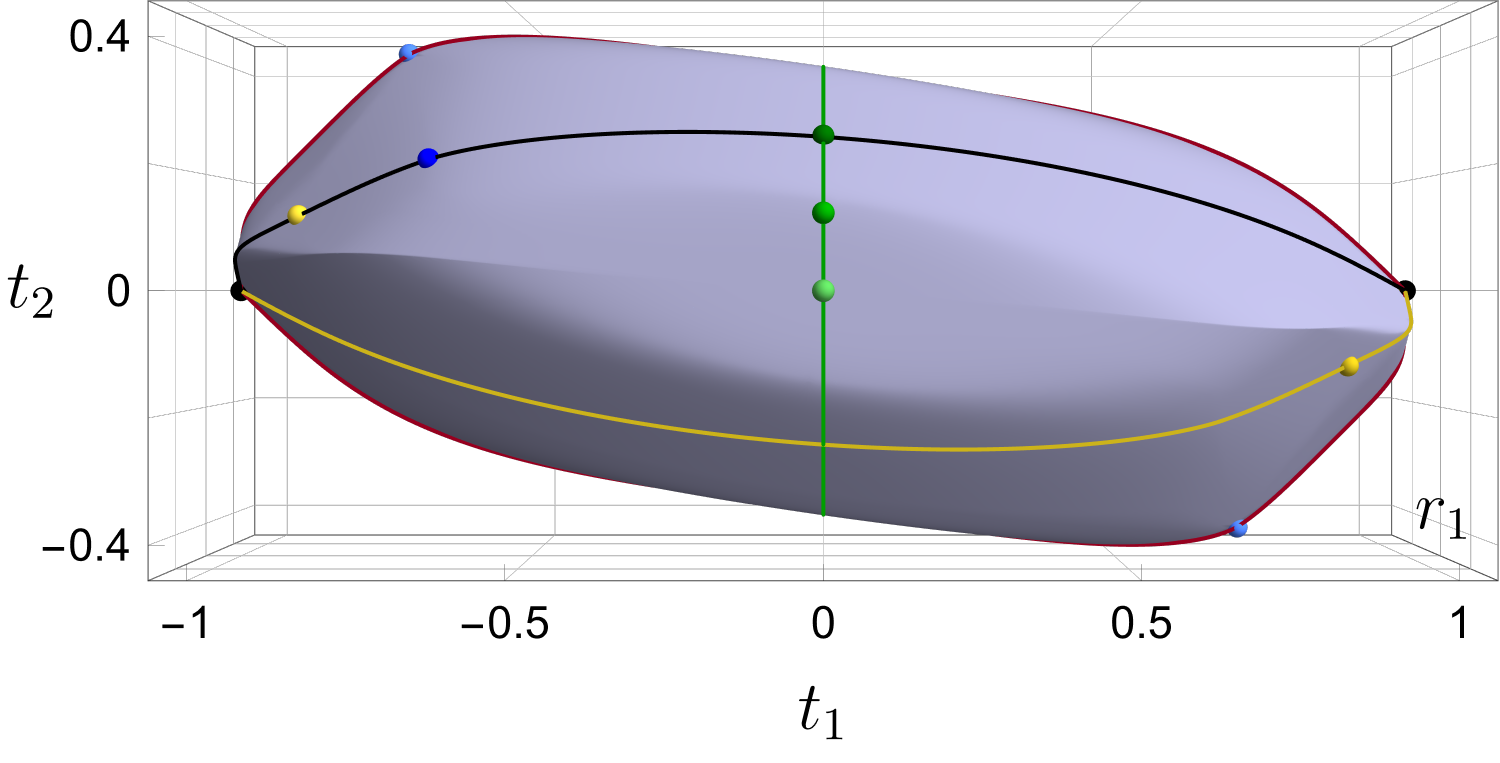}
\end{subfigure}\vspace{0.6cm}
\raisebox{19pt}{
\begin{subfigure}[][][c]
{0.115\textwidth}
\includegraphics[width=.999\textwidth,left,valign=top]{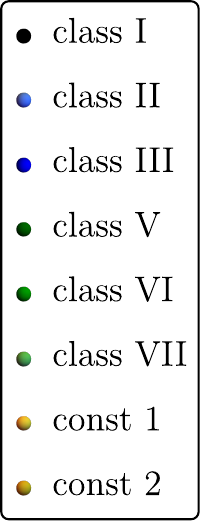}
\end{subfigure}
}
\hfill\vspace{0.2cm}
\begin{subfigure}[][][c]
{0.99\textwidth}\centering
\includegraphics[width=.78\textwidth]{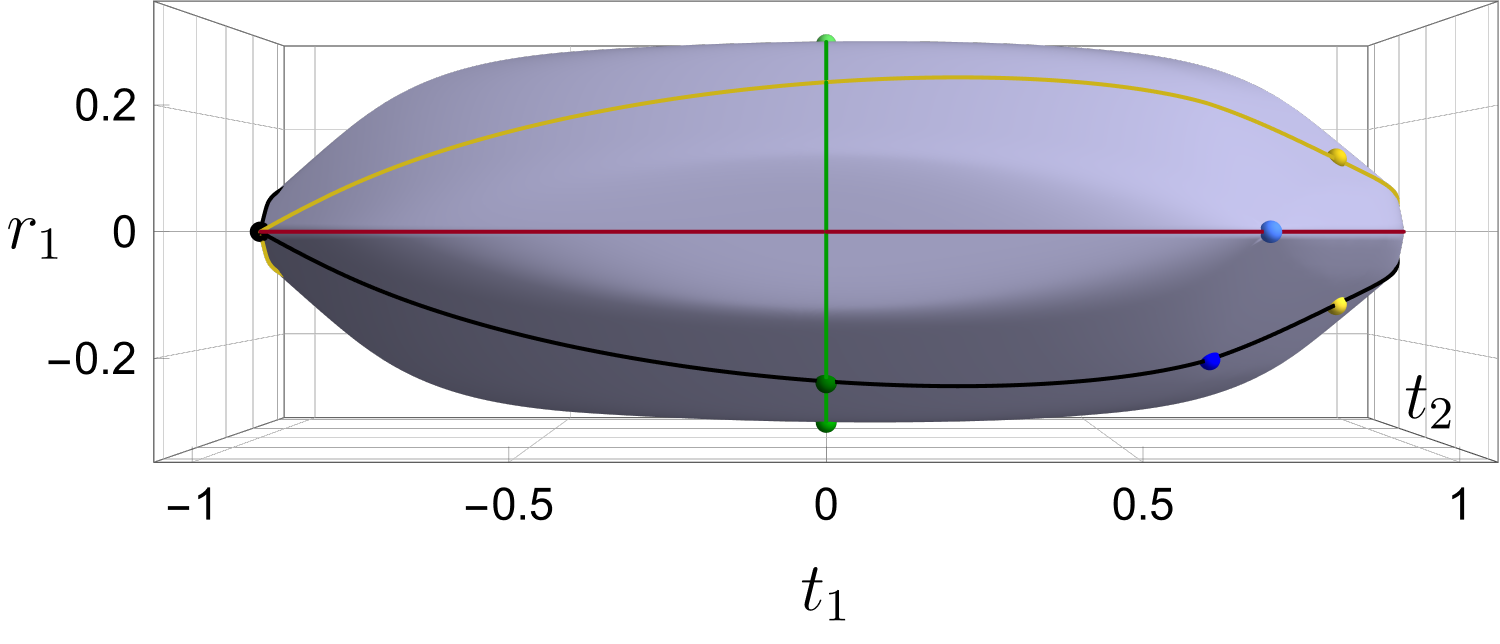}
\end{subfigure}\hfill
\begin{subfigure}[][][c]{0.99\textwidth}\centering
\includegraphics[width=.78\textwidth]{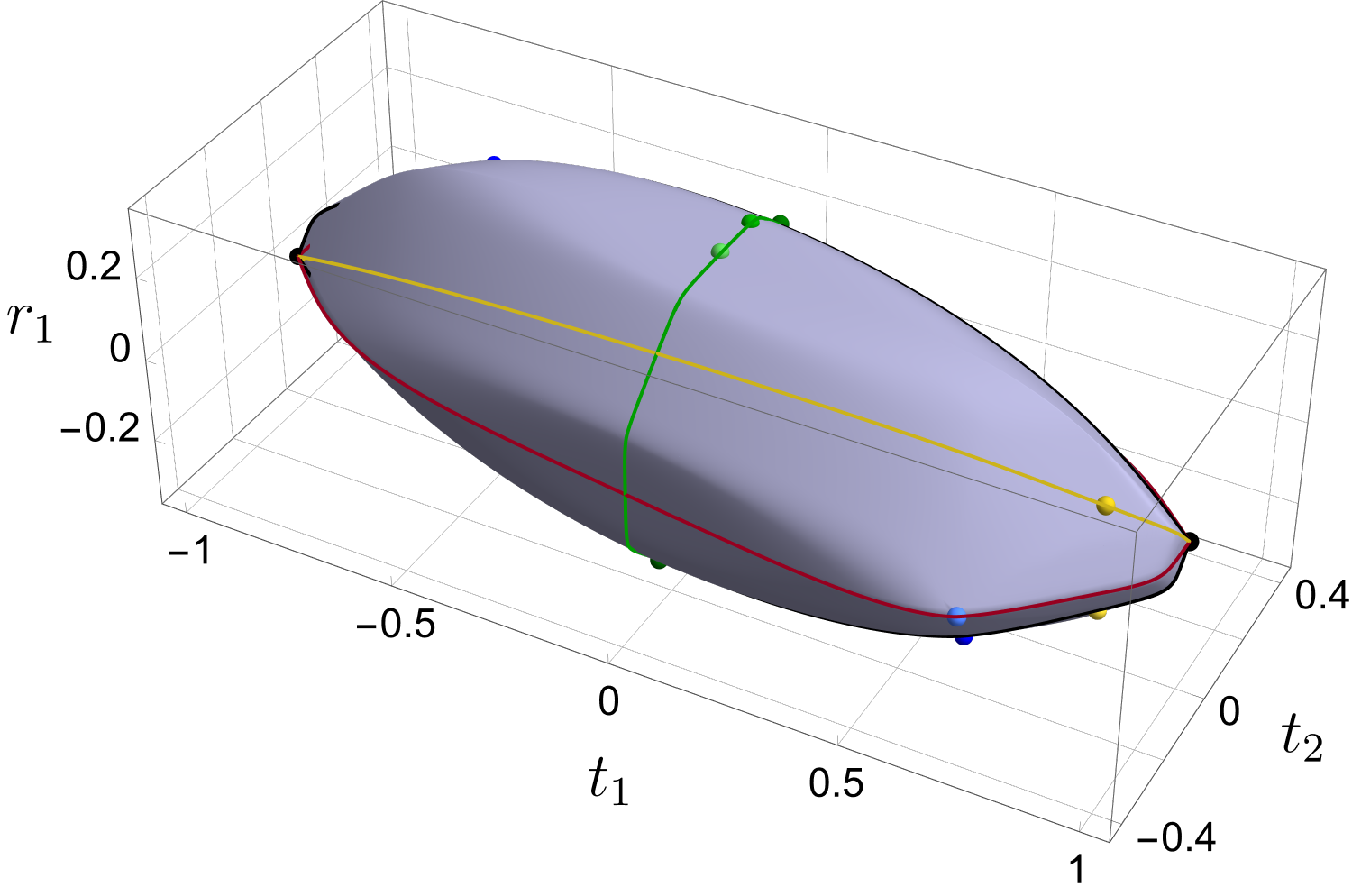}
\end{subfigure}
\caption{U(7) monolith depicting the allowed space of the amplitudes parametrized by $\{t_1,t_2,r_1\}$ (see equations \ref{eq:S matrix def}-\ref{eq:B matrix def}) at the crossing symmetric point $\theta=i\pi/2$. Points at the boundary with exact integrable (table~\ref{table:integrable}) and constant \eqref{eq: def const} amplitudes are highlighted in different colors.
}
\label{fig:N7monos}
\end{figure}

\begin{figure}[th!]
\begin{subfigure}
{0.5\textwidth}
\includegraphics[width=.97\textwidth,left]{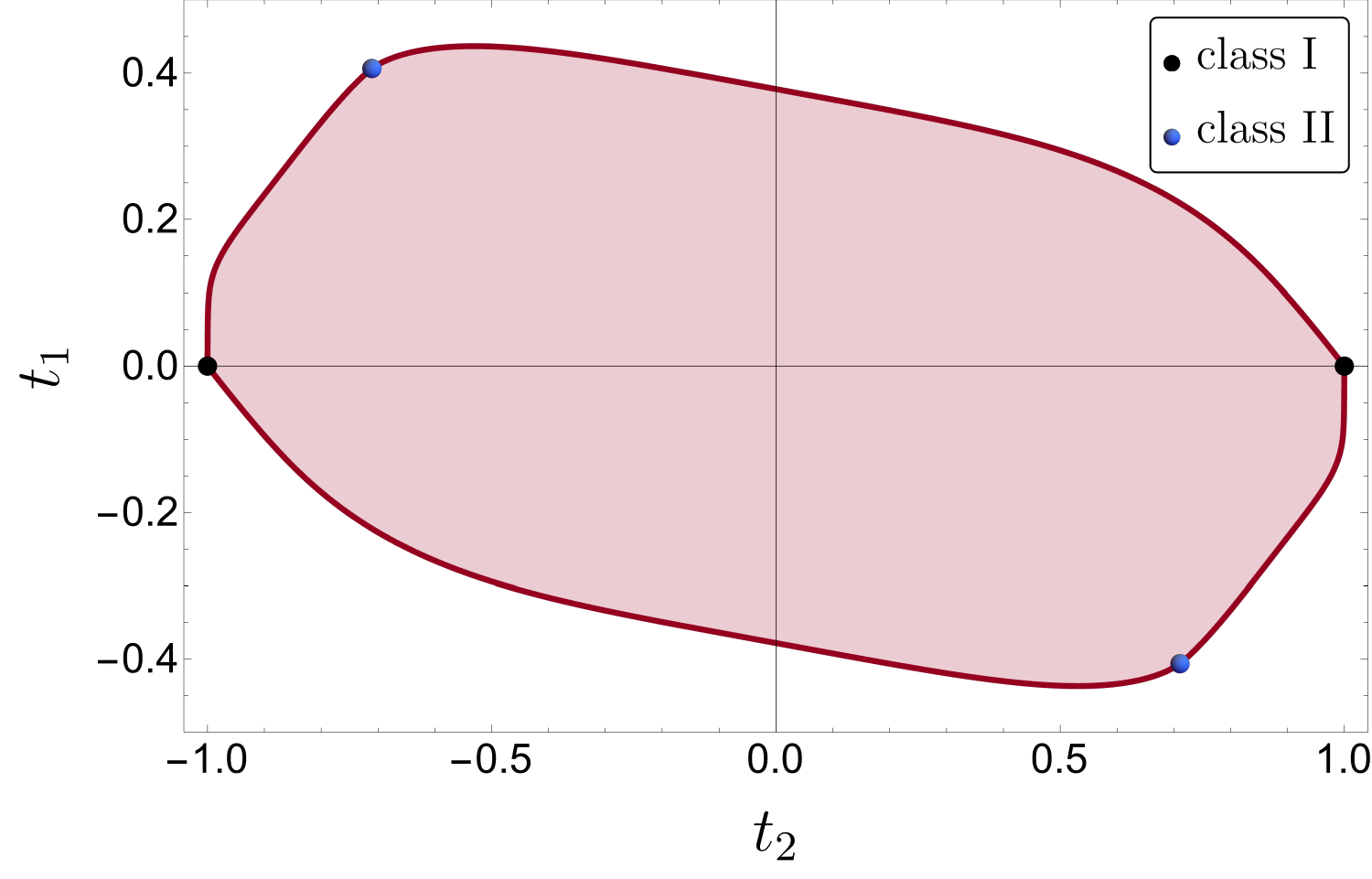}
\caption{$r_1=0$}
\end{subfigure}
\begin{subfigure}
{0.5\textwidth}
\includegraphics[width=.97\textwidth,right]{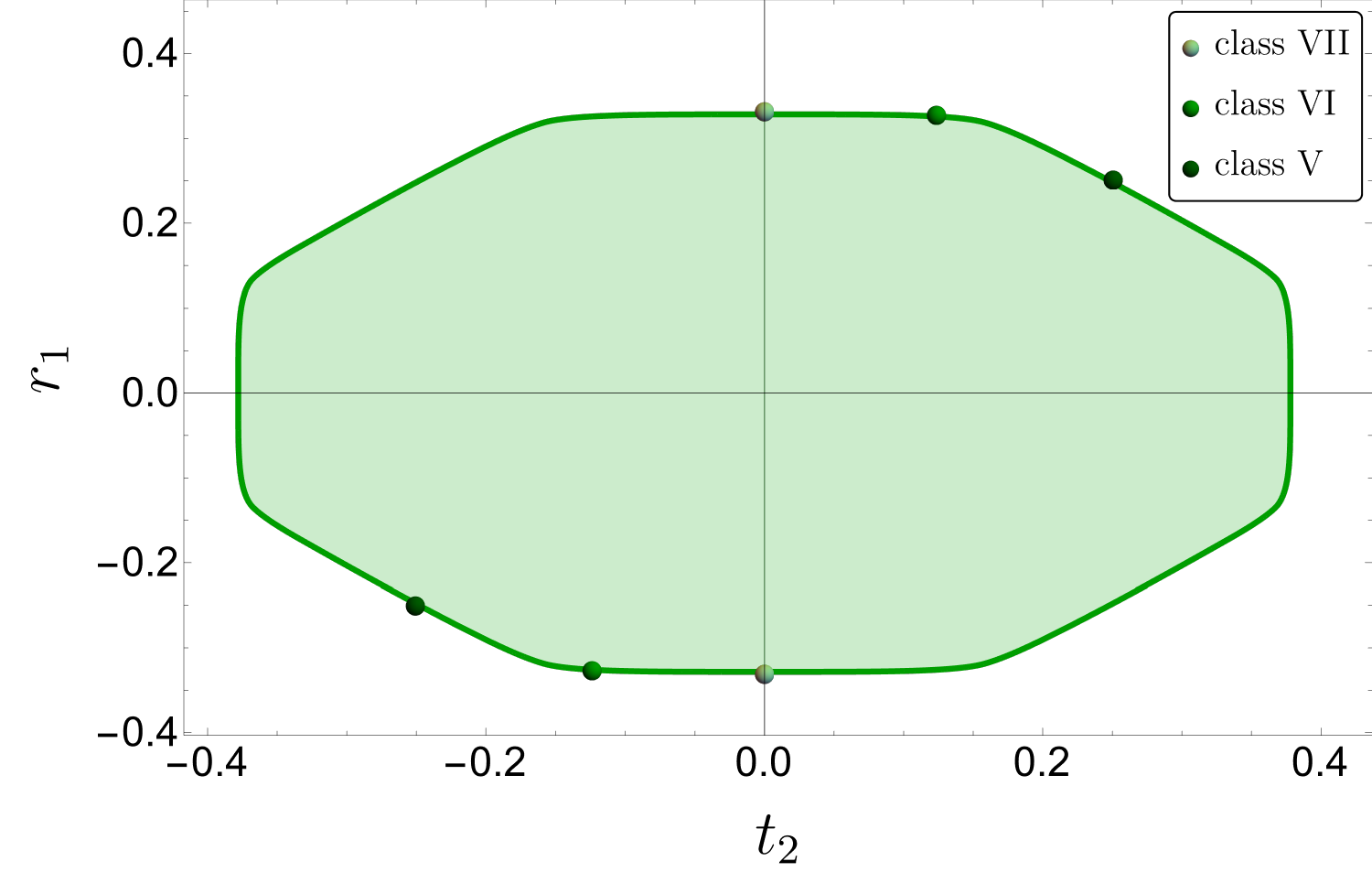}
\caption{$t_1=0$}
\end{subfigure}\vspace{.5cm}
\begin{subfigure}
{0.5\textwidth}
\includegraphics[width=.97\textwidth,left]{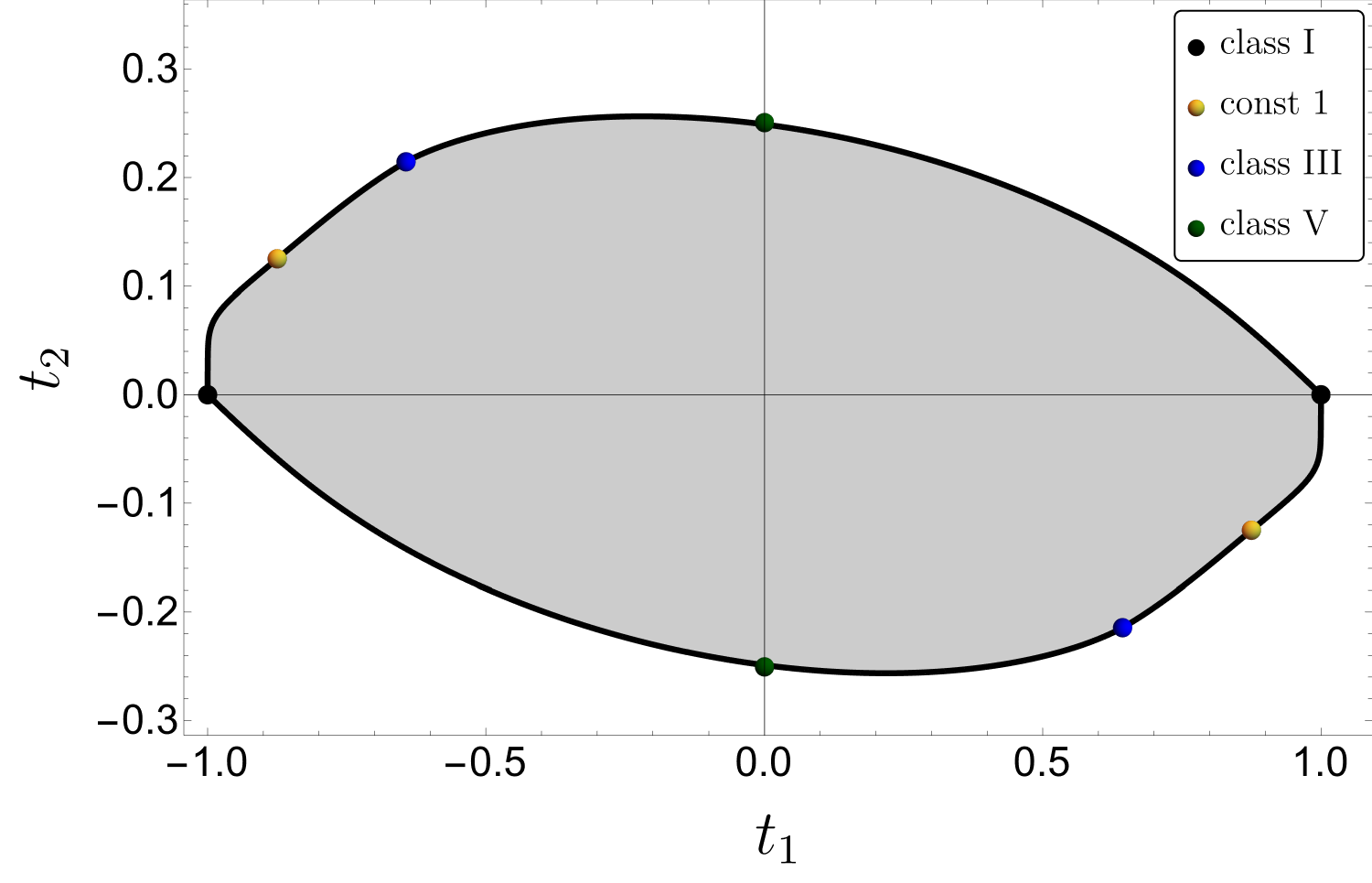}
\caption{$t_2=r_2$}
\end{subfigure}
\begin{subfigure}
{0.5\textwidth}\centering
\includegraphics[width=.97\textwidth,right]{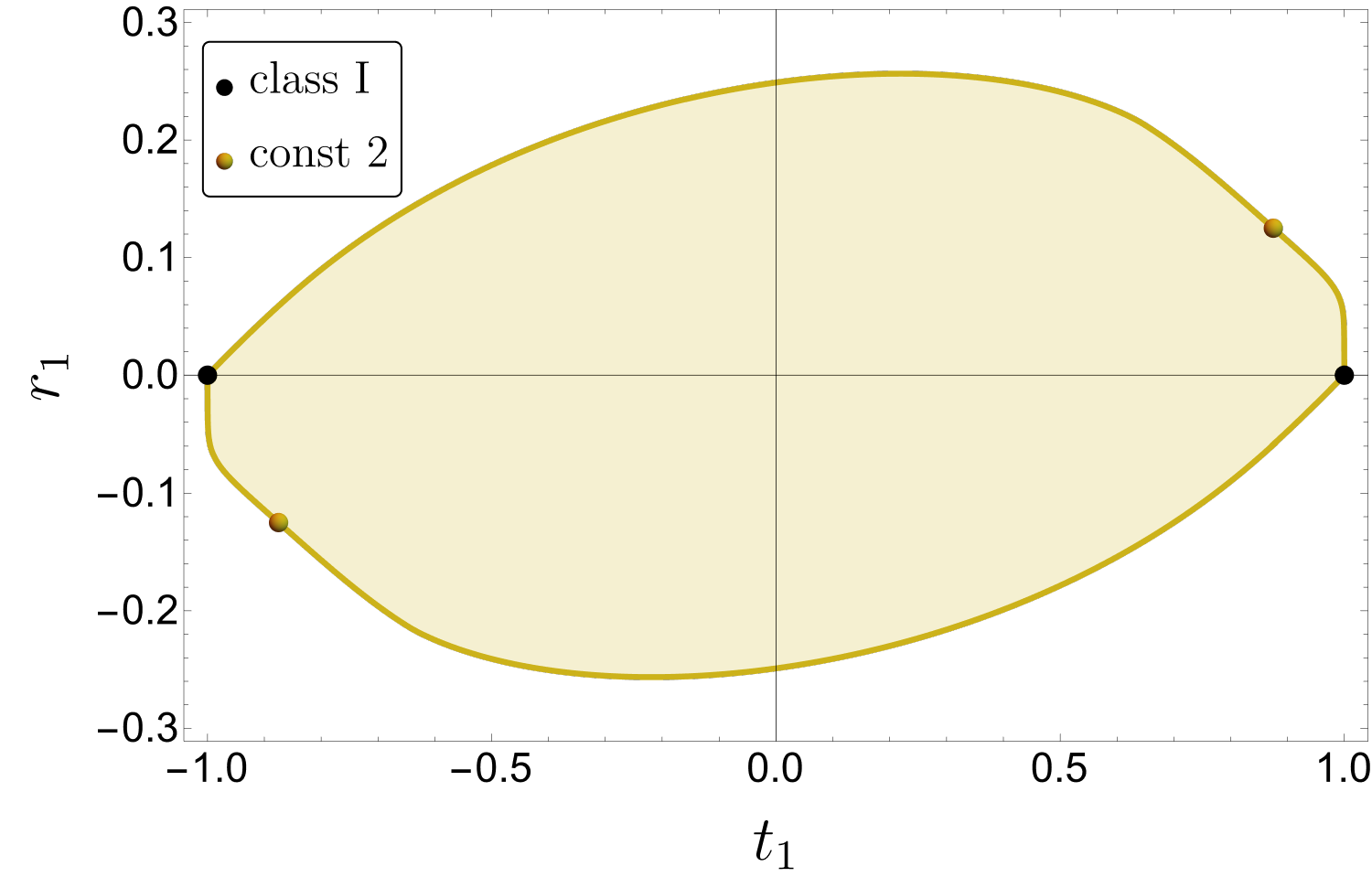}
\caption{$t_2=-r_2$}
\end{subfigure}
\caption{Sections of the $U(7)$ monolith (see solid lines in figure~\ref{fig:N7monos}), obtained from the dual approach code of appendix \ref{app:codes} with \textit{nmax} = $10$, \textit{nintpts} = $30$, \textit{precision} = $100$ and $\sim1000$ angles.}
\label{fig:N7sections}
\end{figure}

In this section we present two- and three-dimensional subsections of the allowed space of two-body amplitudes with global $U(N)$ symmetry. These sections are obtained by placing numerical bounds on the functions $t_1,\,t_2,\,r_1$ (see equations~\ref{eq:S matrix def}-\ref{eq:B matrix def}) at the crossing symmetric point $\theta=i\pi/2$ ($s=2m^2$). Figures~\ref{fig:N2monos}-\ref{fig:N7sections} show these spaces for $N=2$ and $N=7$. The bootstrap implementation utilized to generate these plots is explained in \ref{subsec:numerics} and appendix~\ref{app:codes}. 

The first thing to note from the above figures is that the space of allowed amplitudes is convex. This follows from various properties of the amplitudes: the unitarity condition in physical kinematics, analyticity inside the physical sheet and boundedness at the left hand cut due to crossing. 

Let us now discuss the symmetries of this allowed space. That is, given an extremal amplitude $S_a(\theta)$ saturating the bootstrap constraints,
we want to understand which simple maps lead us to another point at the boundary. Note first that we are allowed to simultaneously change the overall sign of the amplitudes $S_a\rightarrow-S_a$ without spoiling any of the analyticity, crossing and unitarity conditions.\footnote{This is only true if there are no bound states in the theory, as the residue at the bound state pole should have a definite sign.} Other symmetries come from flipping the sign of only some of the $u_i,\, t_i,\, r_i$ components. For instance, sending $r_{1,2}\rightarrow-r_{1,2}$ \footnote{Both components $r_1$ and $r_2$ should be transformed simultaneously due to crossing symmetry.} simply flips the parity even and odd sectors $S_{\text{sing }\pm}\rightarrow S_{\text{sing }\mp},\,S_{\text{adj }\pm}\rightarrow S_{\text{adj }\mp}$. Finally we can also make the following simultaneous change of signs $u_{1,2}\rightarrow-u_{1,2},\,t_{1,2}\rightarrow-t_{1,2}$, which can be thought of as a composition of the latter two transformations.

The spaces of allowed amplitudes presented\footnote{Both figures~\ref{fig:N2monos} and \ref{fig:N7monos} were obtained using the dual approach functions in appendix \ref{app:codes} with $\textit{nmax} = 10$, $\textit{nintpts} = 30$, $\textit{precision} = 100$ and approximately $120,000$ different angles to scan the space.} in figures~\ref{fig:N2monos} and \ref{fig:N7monos} have various geometric features such as sharp edges and vertices. We have identified various integrable points (class I-VII in figures) and constant solutions (const. 1 and 2) for which we can write exact non-perturbative expressions which we review below. 

\paragraph{Integrable amplitudes} These describe theories where there is an infinite number of conserved charges which imply the absence of particle production and factorized scattering. The latter condition is imposed through the Yang-Baxter equations, which tell us we can factorize $3\rightarrow3$ scattering into different sequences of $2\rightarrow2$ processes:
\begin{equation}\label{eq:YBE}
   \includegraphics[width=.4\textwidth,center]{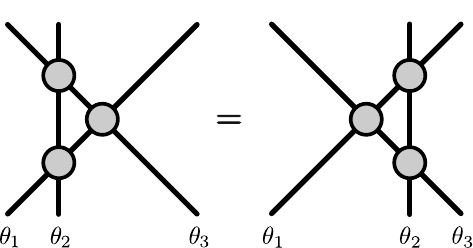}
\end{equation}
When the scattered particles are charged under a global symmetry, these equations are highly constraining and their solutions can be classified. For theories with $U(N)$ symmetry this classification was first performed in \cite{Berg:1977dp}. In table~\ref{table:integrable} we present a revised version of the different classes of integrable amplitudes.\footnote{The classification proposed in table 1 of \cite{Berg:1977dp} has some inaccuracies in the expressions for classes V and VI
and misses two other classes: an extra solution with a free parameter when $N=2$ discussed in subsequent works \cite{Wiegmann:1985jt, Polyakov:1983tt} and what we call class VII in table~\ref{table:integrable}, which to our knowledge has not appeared in the literature before.} In appendix~\ref{app: YB solutions} we review how the different classes arise when solving Yang-Baxter equations. The building block for these amplitudes is the function:
\begin{equation}\label{eq:f_lambda definition}
    f_\lambda(\theta)=\frac{\Gamma(\frac{1}{2}+\frac{\theta}{2\pi i}) \Gamma(\frac{1}{2}+ \frac{\lambda}{2} -\frac{\theta}{2\pi i})}{\Gamma(\frac{1}{2}-\frac{\theta}{2\pi i}) \Gamma(\frac{1}{2}+ \frac{\lambda}{2} +\frac{\theta}{2\pi i})}\,,
\end{equation}
which already captures some of the unitarity and crossing requirements.\footnote{By this we mean that starting from a pole in the complex plane, its images under crossing $\theta\rightarrow i\pi-\theta$ and unitarity $\theta\rightarrow-\theta$ naturally lead to the ratio of gamma functions in \eqref{eq:f_lambda definition}.} 

\begin{table}
\centering
\begin{tabular}{ |c||c|c|c|c|  }
 \hline
 Class & parameter & $t_1(\theta)$ & $t_2(\theta)$ & $r_1(\theta)$ \\
 \hline
  \hline
\rule{0pt}{15pt} I &   & 1 & 0 & 0\\[5pt]
II & $\lambda=\frac{2}{N}$ & $f_\lambda(\theta)$ & $-\, \dfrac{i\pi\lambda}{\tilde\theta}\, t_1(\theta)$ & 0\\[15pt]
III & $\lambda=\frac{1}{N-1}$ & $f_\lambda(\theta)f_\lambda(\tilde\theta)$ & $-\, \dfrac{i\pi\lambda}{\tilde\theta}\, t_1(\theta)$ & $-\, \dfrac{i\pi \lambda}{\theta}\, t_1(\theta)$\\[15pt]
IV & $\lambda=\frac{1}{N+1}$ & $f_\lambda(\theta)f_\lambda(\tilde\theta)\,i\, \text{th} \left(\frac{\theta}{2}-\frac{i\pi}{4}\right)$ & $-\, \dfrac{i\pi\lambda}{\tilde\theta}\, t_1(\theta)$ & $\, \dfrac{i\pi\lambda}{\theta}\, t_1(\theta)$\\[15pt]
V & ch$(\pi \mu)=N$ & 0 & $\dfrac{\sin\mu\theta}{\sin\mu\tilde\theta}\, r_1(\theta)$ & $\displaystyle\prod\limits_{k\in\mathbb Z} \dfrac{f_{1-ik/\mu}(\theta)}{f_{ik/\mu}(\theta)}$\\[15pt]
VI & $e^{\pi\mu}=N$ & 0 & $e^{i\mu\tilde\theta}\,\dfrac{\sin\mu\theta }{\sin\mu\tilde\theta}\, r_1(\theta)$ & $\displaystyle\prod\limits_{k\in\mathbb Z} \dfrac{f_{1-ik/\mu}(\theta)}{f_{ik/\mu}(\theta)}$\\[15pt]
VII & ch$(\pi \mu)=\frac{N}{2}$ & 0 & 0 & $\displaystyle\prod\limits_{k\in\mathbb Z} \dfrac{f_{1-ik/\mu}(\theta)}{f_{ik/\mu}(\theta)}$\\[15pt]
 \hline
  \hline
\rule{0pt}{22pt}   (N=2)$_p$ & $p\,\in [1,\infty)$ & $\dfrac{\text{sh}(\theta/p)}{\text{sh}(\tilde\theta/p)}\,f_1(\theta)\,\xi_p(\theta)$ & $-\dfrac{i\pi}{\tilde\theta}\,t_1(\theta)$ & $-\dfrac{\text{sh}(i\pi/p)}{\text{sh}(\theta/p)}\,t_1(\theta)$ \\[12pt]
 \hline
\end{tabular}
\caption{Classification of integrable amplitudes with $U(N)$ global symmetry. Classes I-VII give solutions to Yang-Baxter equations without free parameters valid for any $N$, whereas the last entry is an extra family of solutions arising when $N=2$. The functions $f_\lambda(\theta)$ and $\xi_p(\theta)$ are respectively defined in \eqref{eq:f_lambda definition} and \eqref{eq:xi_p definition} and we have used the shorthand notations: $\tilde\theta=i\pi-\theta$, sh=$\sinh$, ch=$\cosh$ and th$=\tanh$. These amplitudes are the ``minimal" integrable amplitudes without bound state poles.}
\label{table:integrable}
\end{table}

Now let us discuss the models described by the amplitudes in table~\ref{table:integrable}. Free theories are described by Class I; class II solutions appear in the theories discussed in \cite{Basso:2012bw}, sometimes called massive $CP^{N-1}$ models;\footnote{There is a rich history behind these models, see e.g. \cite{DAdda:1978dle,Abdalla:1981yt,Abdalla:1985nm, Koberle:1982ju, Koberle:1987wc}, which can be seen as the $CP^{N-1}$ model coupled to fermions or as a one parameter deformations of the $O(2N)$ non-linear sigma models. For $N>2$ they are integrable only for special tuning of the parameters \cite{Basso:2012bw}.} and class III coincides with the amplitudes of $O(2N)$ non-linear sigma model \cite{Zamolodchikov:1978xm}. Classes V-VII are solutions to Yang-Baxter equations which are periodic in the real $\theta$ direction, with period $2\pi/\mu$. To the best of our knowledge, no precise identification of physical models has been made for the latter amplitudes (see section~\ref{sec:Discussion} for a discussion on possible connection to loop models and complex conformal theories). Finally, the free parameter solution for $N=2$ describes the deformation of the $O(4)$ non-linear sigma model in \cite{Wiegmann:1985jt,Polyakov:1983tt}. 

Except for class IV for which --because of the $\tanh{\(\frac{\theta-i\pi/2}{2}\)}$ factor in $t_1(\theta)$-- amplitudes vanish at the crossing symmetric point, all of the integrable amplitudes in table~\ref{table:integrable} appear at the boundary of the monolith, i.e. they are \textit{extremal} amplitudes. These points are highlighted in different colors in figures~\ref{fig:N2monos}-\ref{fig:N7sections}. Note that the transformation $S_a\rightarrow -S_a$ gives an equally good solution to the Yang Baxter equations, so that for each class we have two integrable points on the monolith. Moreover, classes I-III are vertices of the allowed regions for any $N$ and class VI for $N=2$.

\paragraph{Constant solutions} Another set of extremal amplitudes for which we have exact solutions is given by constant amplitudes which, in contrast to free theory, do not saturate the unitarity constraint in all channels. In the notation of \eqref{eq:S matrix components vector form} we have
\begin{equation}\label{eq: def const}
    \mathbf S_{\pm\text{ const 1}}=\pm \begin{pmatrix} 1\\ \frac{1-N}{1+N} \\ -1 \\ -1 \\  \frac{1-N}{1+N}  \\ -1  \end{pmatrix}\,,\;\qquad\qquad \mathbf S_{\pm\text{ const 2}}=\pm \begin{pmatrix} -1\\ -1 \\ 1 \\ \frac{1-N}{1+N} \\  \frac{1-N}{1+N}  \\ -1  \end{pmatrix} \,,
\end{equation}
where the two solutions are related by the exchange of parity even and odd sectors. This type of solutions were called yellow points in \cite{Cordova:2019lot} and are highlighted in different shades of that color in figures~\ref{fig:N2monos}-\ref{fig:N7sections}. Since they are solutions to both the simplest primal and dual bootstrap problems, that is giving $S_a=c_a$ and $K_a(\theta)=i n_a/\cosh\theta$ as ansatze, these points sit in the middle of flat faces (see e.g. figure 4 in \cite{Cordova:2019lot}).

\paragraph{Unitarity saturation} We observe that all extremal amplitudes --except for the constant solutions \eqref{eq: def const}-- saturate the unitarity constraint $|S_a(\theta\geq0)|^2\leq1$. 
This is a common (albeit unwelcome) feature in S-matrix bootstrap explorations, as non-integrable theories should allow for multiparticle processes $S_{2\rightarrow n>2}\neq0$. Indeed, in the absence of multiparticle data the algorithm that maximizes a given functional will produce amplitudes which are as large as possible at the boundary of the physical sheet. Nevertheless, low energy observables such as the bounds discussed here are virtually unaffected by possible particle production at higher energies. Our claim is that the amplitudes found here still provide a good low-energy approximation to physical ones.\footnote{For a discussion on how adding an arbitrary function for particle production at a given energy threshold modifies both the bounds and the amplitudes' analytic structure see e.g \cite{Cordova:2018uop,Antunes:2023irg}.}

\paragraph{Sections and global symmetry} In figures~\ref{fig:N2sections} and~\ref{fig:N7sections} we show different sections of the $U(N)$ monoliths. The first one labeled by (a) is the section where $r_{1,2}(i\pi/2)=0$. What we see is that the extremal amplitudes obey $r_{1,2}(\theta)=0$ for all energies, so that there is no backward scattering. There are two (four, taking into account the $S_a\rightarrow-S_a$ symmetry) vertices corresponding to integrable theories: class I, i.e. free theory, and class II. In section (b) we find that extremal amplitudes obey $t_1(\theta)=u_1(\theta)=0$, so that the transmission part of the amplitude vanishes. It is in this section that we find the periodic integrable amplitudes (classes V-VII) and class VI sits at a vertex for $N=2$.  The extremal amplitudes in section (c) satisfy $t_2(\theta)=r_2(\theta)$ and $u_1(\theta)=t_1(\theta)$ (as well as $u_2(\theta)=r_1(\theta)$ from crossing). What this means is that these amplitudes do not differentiate particles from antiparticles and the global symmetry is enhanced to $O(2N)$. The section indeed reproduces the results of \cite{Cordova:2019lot}, where the exact amplitudes highlighted with points are: free theory, the first constant solution in \eqref{eq: def const}, the $O(2N)$ non-linear sigma model and the $O(2N)$ ``periodic Yang-Baxter". Finally section (d) shows the same amplitudes as in (c) except for the change of sign: $r_{1,2}\rightarrow-r_{1,2}$, which for non-trivial amplitudes spoils integrability. 

\begin{figure}[!ht]
    \centering
    \includegraphics[width=0.99\linewidth]{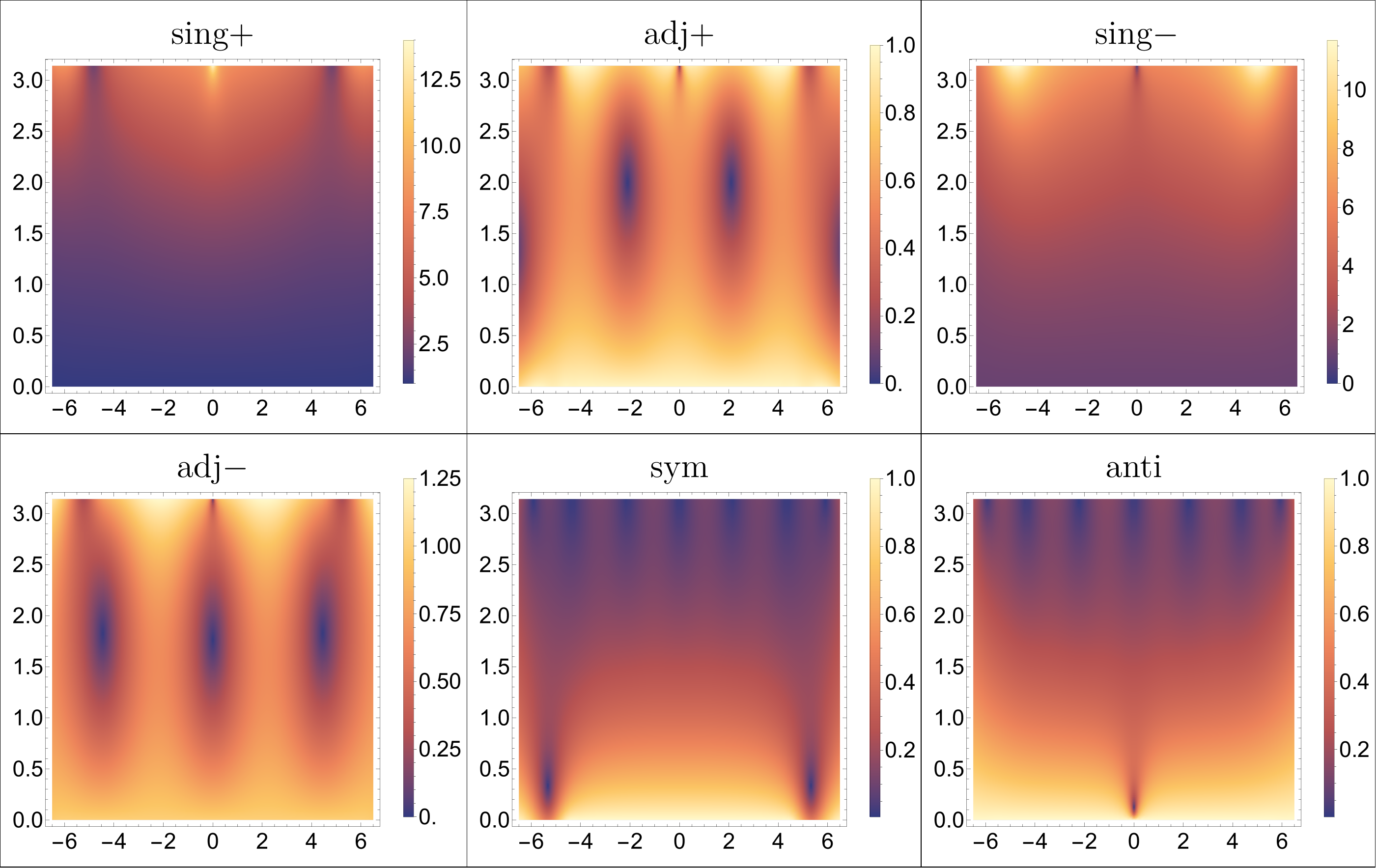}
    \caption{Analytic structure of the amplitudes in the complex $\theta$ plane for a generic point on the U(7) monolith close to constant 1 solution \eqref{eq: def const}. The plots show an array of zeros signaling resonances in different channels. For these we used the primal bootstrap code with  $\textit{nmax}=60$, $\textit{ngrid=120}$.}
    \label{fig:analyticcomplextheta}
\end{figure}

\begin{figure}[!h]
    \centering
    \includegraphics[width=0.99\linewidth]{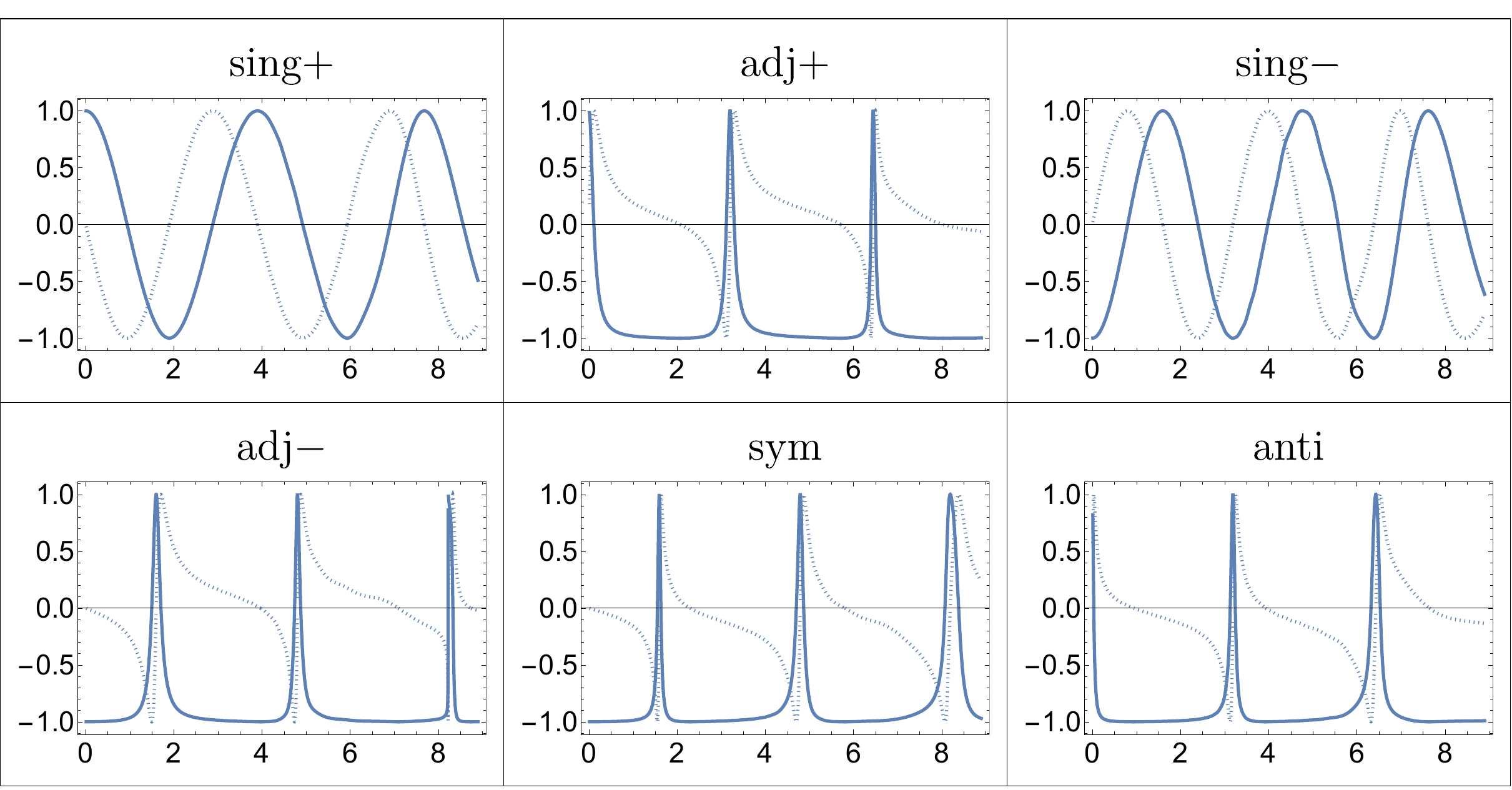}
    \caption{Real (solid) and imaginary (dashed) parts of the amplitudes $S_a(\theta)$ in physical kinematics $\theta\in\mathbb R$ for an extremal point close to constant 1 \eqref{eq: def const} with $N=1000$. These numerical amplitudes show the emergent periodicity present in most of the monolith. To generate these plots with the dual code we used $\textit{nmax}=130$, $\textit{nintpts}=350$, $\textit{precision}=350$.}
    \label{fig:periodicity}
\end{figure}

\paragraph{Resonances and periodicity} Similar to the space of $O(N)$ amplitudes in \cite{Cordova:2018uop,Cordova:2019lot}, the extremal amplitudes generically exhibit an infinite number of resonances in the physical Riemann sheet. These can be detected as zeros in the complex rapidity plane with $0\leq\text{Im}(\theta)\leq\pi$. An example is presented in figure~\ref{fig:analyticcomplextheta} for a point on the boundary of the $U(7)$ monolith, close to the constant 1 solution. We also observe that these resonances arrange themselves in a way such that the amplitude is periodic. We show this periodic behavior in physical kinematics in figure~\ref{fig:periodicity} for an extremal amplitude with $N=1000$.\footnote{The periods generically decrease with $N$, so we chose a large number of flavors to show the periodicity up to ``low" energies $\theta\approx8$ ($s\approx 3000m^2$) where the numerical amplitudes are more reliable.} While this periodicity in (a parametrization of the) energy might seem strange from a physical point of view,  it is present in the vast majority of extremal amplitudes with group-like symmetries. More on this point and the possible connection to complex conformal field theories are discussed in the next section.

\section{Discussion}\label{sec:Discussion}

In this work we have begun to chart the space of amplitudes with a global $U(N)$ symmetry for massive Quantum Field Theories in two spacetime dimensions. We focused on the two-to-two scattering of particles in the fundamental/antifundamental representation of U(N) without bound states. We observed integrable models emerging at special points of the boundary of the allowed regions. Most of these integrable amplitudes were written down in \cite{Berg:1977dp}. Table~\ref{table:integrable} summarizes the integrable solutions compatible with this symmetry, including two that were missing from the original classification in \cite{Berg:1977dp}.

The first new solution, labeled (N=2)$_p$ in table~\ref{table:integrable}, can be viewed as a one-parameter deformation of the $O(4)$ nonlinear sigma model and was previously discussed in, e.g., \cite{Wiegmann:1985jt,Polyakov:1983tt,Basso:2012bw}. The second, Class VII in the same table, has no free parameters and represents another Yang-Baxter solution characterized by periodic behavior in real rapidity. To the best of our knowledge, this solution has not previously appeared in the literature.

One of the more puzzling aspects we encountered is the ubiquity of extremal amplitudes with periodic behavior in real rapidity, as has been previously observed for theories with conventional group symmetries \cite{Cordova:2019lot,Bercini:2019vme}. It would be very valuable to understand analytically why the bootstrap optimization problem leads to such space of solutions and why the presence of non-invertible symmetries seems to explore a different space of solutions \cite{Copetti:2024rqj, Copetti:2024dcz}. While particle production will ultimately break this periodicity, we believe that the amplitudes we found remain a good approximation of the physical ones at low energies.

An exception is given by integrable models where multiparticle processes are factorized into a sequence of $2\rightarrow2$ amplitudes and particle production is absent. Classes V-VII in table~\ref{table:integrable} belong to this category, characterized by both integrability and periodic behavior. Although the physical realizations of these solutions have not been fully established, we expect them to be related to deformations of complex conformal theories. For the $O(N)$ periodic Yang-Baxter solution in \cite{Hortacsu:1979pu,Cordova:2018uop,Cordova:2019lot} such a connection was proposed in \cite{Gorbenko:2020xya}. 

We briefly review their argument here. The conformal theories are given by the continuum limit of critical $O(N)$ loop models \cite{Domany:1981fg}, which have two (real) fixed points for $-2\leq N\leq2$. From the higher temperature fixed point (also called dilute model) there is a massive deformation described by the integrable $O(|N|\leq2)$ amplitude of Zamolodchikov \cite{Zamolodchikov:1990dg}. As we approach  $N=2$, the two fixed points merge and become complex for $N>2$. What the authors of \cite{Gorbenko:2020xya} observed is that the $O(N)$ periodic Yang-Baxter solution, naturally defined for $N>2$, describes the expected walking behavior for the RG flow passing between the two complex CFTs. That is, for a large range of energies, quantities such as the central charge look constant. The walking behavior is more prominent for values of $N$ close to the merging point of CFTs, where the complex CFTs are close to the real line. 

\begin{figure}[t]
\begin{subfigure}
{0.49\textwidth}
\includegraphics[width=.99\textwidth,left]{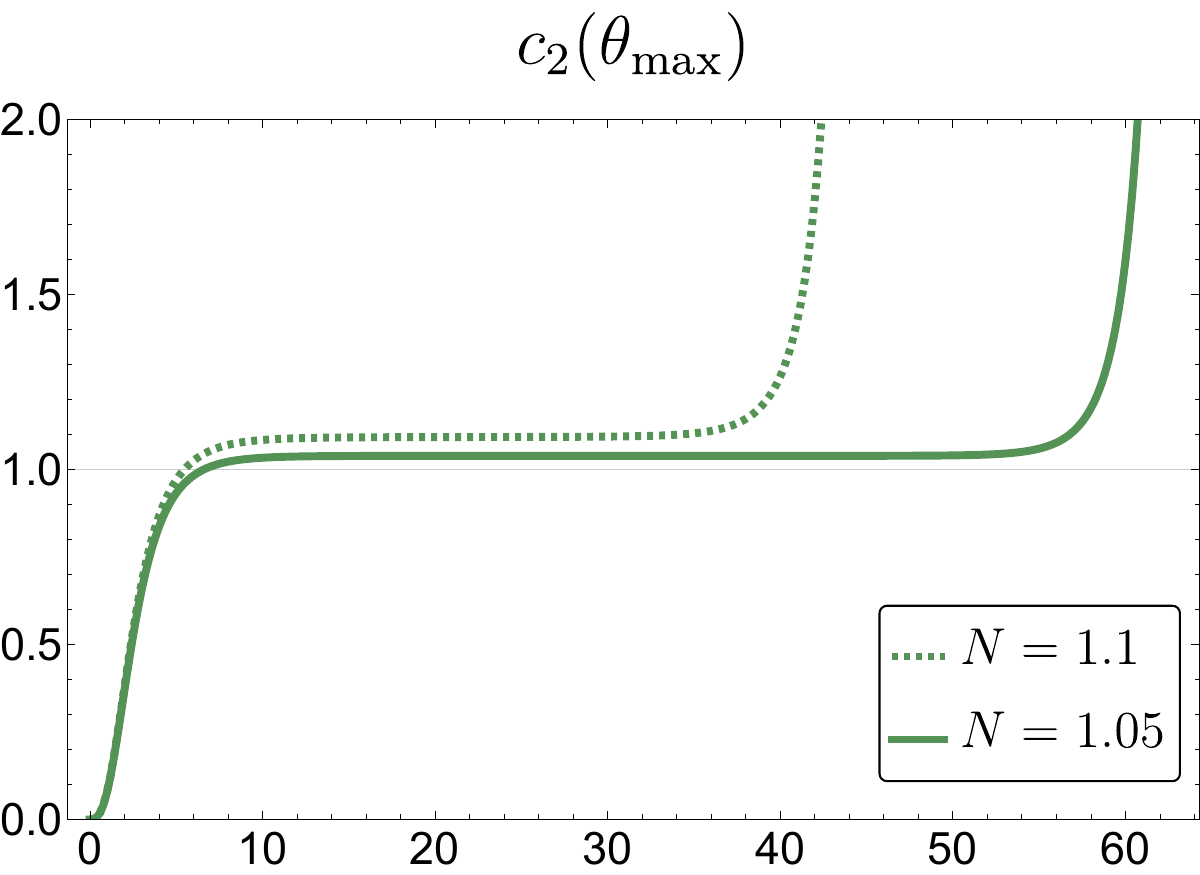}
\caption{Class V}
\label{fig:walkingV}
\end{subfigure}
\begin{subfigure}
{0.49\textwidth}
\includegraphics[width=.955\textwidth,right]{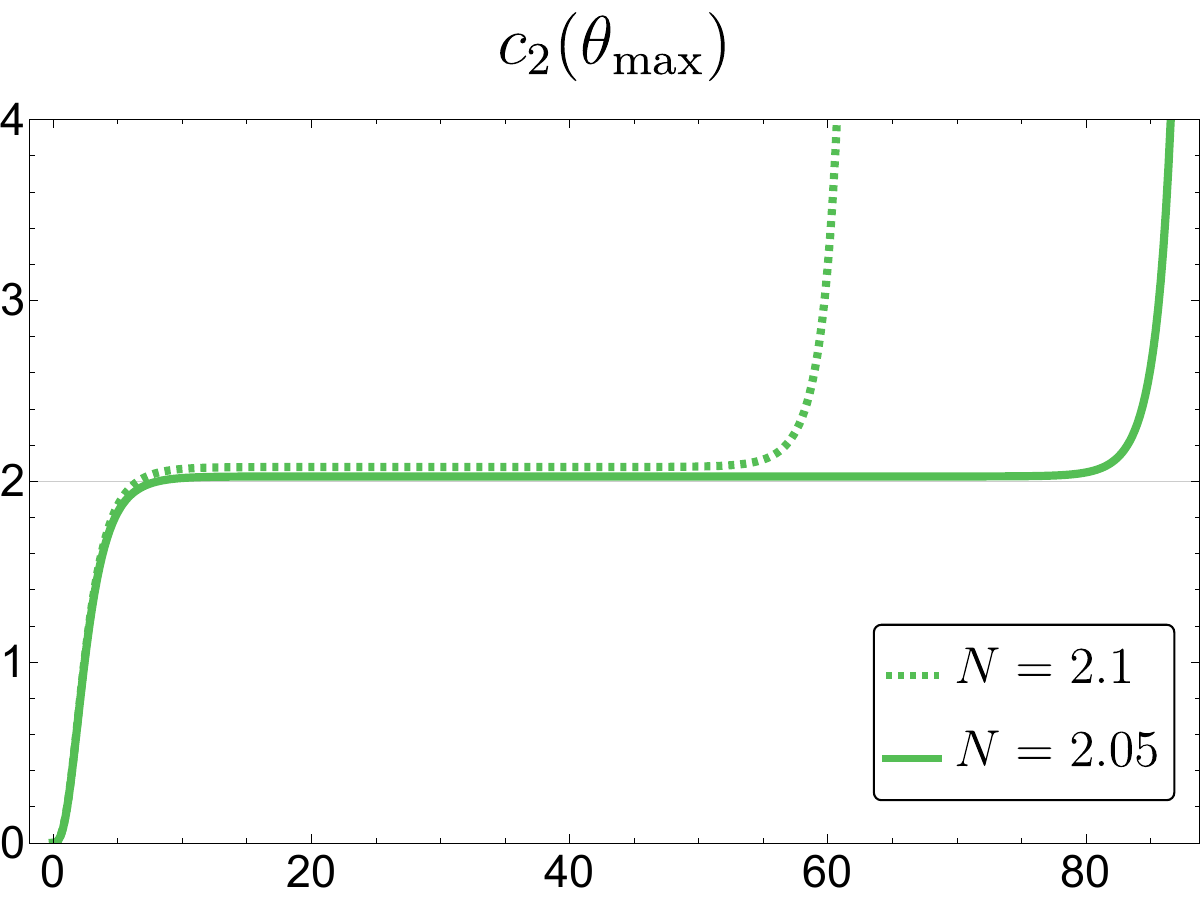}
\caption{Class VII}
\label{fig:walkingVII}
\end{subfigure}
\caption{Walking behavior of the central charge for $U(N)$ periodic integrable amplitudes.}
\label{fig:walking}
\end{figure}

What we propose here is that the $U(N)$ periodic amplitudes studied in this work should describe similar walking behaviors. We test this with the integrable amplitudes in classes V-VII, where we can  easily compute the two-particle contribution to the UV central charge $c_2$ using the two particle form factors of the trace of the stress tensor $\Theta$ as explained in appendix~\ref{app:form factor expressions}. The c-sum rule integral reads
\begin{equation}
    c_2(\theta_\text{max})= 3\int\limits_0^{\theta_\text{max}} d\theta\, \frac{|F_2^\Theta(\theta)|^2}{16\cosh^4\frac{\theta}{2}}\,.
    \label{eq:two particle contribution to central charge}
\end{equation}
Here we are performing the integral up to some maximum rapidity $\theta_\text{max}$ to test the walking behavior.\footnote{The full central charge is given by an integral over all energies, i.e. $\theta_\text{max}\rightarrow\infty$, which often diverges for these extremal amplitudes (see \cite{Cordova:2023wjp} for the $O(N)$ case). What we are showing here is the approximately constant behavior for a large range of energies.} As we show in figure~\ref{fig:walkingV}, we find that classes V-VI walk for $N\gtrsim1$  around $c=1$.\footnote{We are quoting the values of $c_2$ for the integrable amplitudes with an overall minus sign with respect to table~\ref{table:integrable}. The ones with a plus sign walk for the same values of $N$ but have larger central charge.} For class V this behavior is expected, as it corresponds to the $O(2N)$ periodic Yang-Baxter solution and for $N=1$ coincides with the $O(2)$ NLSM for which $c=1$. The central charge value for class VI is the same as in class V, as the amplitudes and form factors coincide for $N=1$. 

Concerning class VII in figure~\ref{fig:walkingVII}, we see walking behavior for $N\gtrsim2$ around $c=2$. A natural question to ask is what is the corresponding UV conformal theory for $N\leq2$? A candidate theory is given by the critical $U(N)$ loop model studied in \cite{Roux:2024ubh}. In this model, non-intersecting oriented loops are considered, where all neighboring loops have opposite orientations. Following Zamolodchikov's logic in \cite{Zamolodchikov:1990dg}, we expect the amplitude to contain only the tensor structures associated to $r_{1,2}$ in \eqref{eq:B matrix def}, which is indeed the case in class VII. A second hint is that this loop model is conjectured to have the same phase diagram as the $O(N)$ one, so that there is a merging of fixed points at $N=2$. 

There is a discrepancy, however, with the expected central charge at $N=2$. Our model exhibits walking behavior around $c=2$ whereas the fixed point studied in \cite{Roux:2024ubh} has $c=1$ for this value of $N$. We should note that the actual symmetry of the model in  \cite{Roux:2024ubh} is $PSU(N)$, since it can be constructed from a spin chain that alternates between fundamental and antifundamental degrees of freedom at each site. 
With this global symmetry, fundamental excitations are not allowed. In an analogy with the $CP^{N-1}$ model, one expects only meson-like excitations. Nevertheless, if an appropriate term to the $CP^{N-1}$ Lagrangian such as a minimally coupled Dirac fermion with quartic interaction is added, the fundamental excitations become ``liberated". Indeed, this modified model gives rise to the integrable amplitudes of class II and (N=2)$_p$ for a particular choice of parameters \cite{Basso:2012bw}. Moreover, since it interpolates between the $O(2N)$ NLSM and $CP^{N-1}$ + free boson, the UV central charge is $c=2N-1$ (i.e. one unit larger than for $CP^{N-1}$). It is possible that a similar mechanism could be at play in the $PSU(N)$ loop models and explain the discrepancy in the central charges. It would be very interesting to understand better this connection, find other loop models with full $U(N)$ global symmetry, and identify a candidate loop model for class VI.

In this work we have added a new piece to the map of two-dimensional QFTs.
As with other S-matrix bootstrap studies,  generic features of the allowed amplitudes point to interesting behaviors that go beyond conventional expectations. This suggests two paths forward: either formulate new criteria to restrict the space to conventional theories or embrace the exploration of the novel features. In either case, continued investigation will undoubtedly deepen our understanding of non-perturbative quantum field theory.

\section*{Acknowledgements}
The authors would like to thank Benjamin Basso, Vasco Gon\c{c}alves and Emilio Trevisani for useful discussions. Centro de Fisica do Porto is partially funded by Funda\c{c}\~ao para a Ciencia e
a Tecnologia (FCT) under the grant UID04650-FCUP. R.R. is supported by FCT under the grant 2024.04788.BD, by FCT grant 2024.00230.CERN and HORIZON-MSCA-2023-SE-01-101182937-HeI.

\newpage
\appendix


\section{Solutions to Yang-Baxter equations}\label{app: YB solutions}
\begin{figure}
    \centering
    \begin{tikzpicture}[thick, scale=2]
        \draw[decoration={markings, mark=at position 0.55 with {\arrow{stealth}},mark=at position 0.08 with {\arrow{stealth}},mark=at position 0.96 with {\arrow{stealth}}},
        postaction={decorate}] (-1.5,0) -- (-1.5,2);
        \draw[decoration={markings, mark=at position 0.55 with {\arrow{stealth}},mark=at position 0.08 with {\arrow{stealth}},mark=at position 0.96 with {\arrow{stealth}}},
        postaction={decorate}] (-1.86,0.186) -- (-0.14,1.21);
        \draw[decoration={markings, mark=at position 0.55 with {\arrow{stealth}},mark=at position 0.08 with {\arrow{stealth}},mark=at position 0.96 with {\arrow{stealth}}},
        postaction={decorate}] (-1.86,1.82) -- (-0.14,0.78);
        \draw[fill=black!10] (-1.5,1.6) circle (5pt);
        \draw[fill=black!10] (-0.5,1) circle (5pt);
        \draw[fill=black!10] (-1.5,0.4) circle (5pt);
        \node[font=\fontsize{9}{9}] at (-1.5,1.6) {$\textrm{F}$};
        \node[font=\fontsize{9}{9}] at (-0.5,1) {$\textrm{F}$};
        \node[font=\fontsize{9}{9}] at (-1.5,0.4) {$\textrm{S}$};

        \node at (-1.86,0.03) {$\theta_1$};
        \node at (-1.5,-0.15) {$\theta_2$};
        \node at (-0.14,0.62) {$\theta_3$};

        \node at (-1.86,1.96) {$\theta_3$};
        \node at (-1.5,2.15) {$\theta_2$};
        \node at (-0.14,1.34) {$\theta_1$};
        
        \node at (0.25,1) {=};
        
        \draw[decoration={markings, mark=at position 0.55 with {\arrow{stealth}},mark=at position 0.08 with {\arrow{stealth}},mark=at position 0.96 with {\arrow{stealth}}},
        postaction={decorate}] (2,0) -- (2,2);
        \draw[decoration={markings, mark=at position 0.55 with {\arrow{stealth reversed}},mark=at position 0.08 with {\arrow{stealth reversed}},mark=at position 0.96 with {\arrow{stealth reversed}}},
        postaction={decorate}] (2.36,0.186) -- (0.64,1.21);
        \draw[decoration={markings, mark=at position 0.55 with {\arrow{stealth reversed}},mark=at position 0.08 with {\arrow{stealth reversed}},mark=at position 0.96 with {\arrow{stealth reversed}}},
        postaction={decorate}] (2.36,1.82) --(0.64,0.78);
        \draw[fill=black!10] (2,1.6) circle (5pt);
        \draw[fill=black!10] (1,1) circle (5pt);
        \draw[fill=black!10] (2,0.4) circle (5pt);
        \node[font=\fontsize{9}{9}] at (2,1.6) {$\textrm{S}$};
        \node[font=\fontsize{9}{9}] at (1,1) {$\textrm{F}$};
        \node[font=\fontsize{9}{9}] at (2,0.4) {$\textrm{F}$};

        \node at (2.36,0.03) {$\theta_3$};
        \node at (2,-0.15) {$\theta_2$};
        \node at (0.64,0.62) {$\theta_1$};

        \node at (0.64,1.34) {$\theta_3$};
        \node at (2,2.15) {$\theta_2$};
        \node at (2.36,1.96) {$\theta_1$};
    \end{tikzpicture}

    \vskip 0.5cm
    
    \begin{tikzpicture}[thick, scale=2]
        \draw[decoration={markings, mark=at position 0.55 with {\arrow{stealth}},mark=at position 0.08 with {\arrow{stealth}},mark=at position 0.96 with {\arrow{stealth reversed}}},
        postaction={decorate}] (-4,-2.5) -- (-4,-0.5);
        \draw[decoration={markings, mark=at position 0.55 with {\arrow{stealth}},mark=at position 0.08 with {\arrow{stealth}},mark=at position 0.96 with {\arrow{stealth}}},
        postaction={decorate}] (-4.36,-2.314) -- (-2.64,-1.29);
        \draw[decoration={markings, mark=at position 0.55 with {\arrow{stealth}},mark=at position 0.08 with {\arrow{stealth reversed}},mark=at position 0.96 with {\arrow{stealth}}},
        postaction={decorate}] (-4.36,-0.68) --(-2.64,-1.72);
        \draw[fill=black!10] (-4,-0.9) circle (5pt);
        \draw[fill=black!10] (-3,-1.5) circle (5pt);
        \draw[fill=black!10] (-4,-2.1) circle (5pt);
        \node[font=\fontsize{9}{9}] at (-4,-0.9) {$\textrm{B}$};
        \node[font=\fontsize{9}{9}] at (-3,-1.5) {$\textrm{F}$};
        \node[font=\fontsize{9}{9}] at (-4,-2.1) {$\textrm{S}$};
        \node at (-2.2,-1.5) {=};        
        \draw[decoration={markings, mark=at position 0.55 with {\arrow{stealth}},mark=at position 0.08 with {\arrow{stealth}},mark=at position 0.96 with {\arrow{stealth reversed}}},
        postaction={decorate}] (-0.4,-0.5) -- (-0.4,-2.5);
        \draw[decoration={markings, mark=at position 0.55 with {\arrow{stealth}},mark=at position 0.08 with {\arrow{stealth reversed}},mark=at position 0.96 with {\arrow{stealth}}},
        postaction={decorate}] (-0.04,-2.314) -- (-1.76,-1.29);
        \draw[decoration={markings, mark=at position 0.55 with {\arrow{stealth reversed}},mark=at position 0.08 with {\arrow{stealth reversed}},mark=at position 0.96 with {\arrow{stealth reversed}}},
        postaction={decorate}] (0.04,-0.68) -- (-1.76,-1.72);
        \draw[fill=black!10] (-0.4,-0.9) circle (5pt);
        \draw[fill=black!10] (-1.4,-1.5) circle (5pt);
        \draw[fill=black!10] (-0.4,-2.1) circle (5pt);
        \node[font=\fontsize{9}{9}] at (-0.4,-0.9) {$\textrm{F}$};
        \node[font=\fontsize{9}{9}] at (-1.4,-1.5) {$\textrm{S}$};
        \node[font=\fontsize{9}{9}] at (-0.4,-2.1) {$\textrm{B}$};
        \node at (0.4,-1.5) {+};
        \draw[decoration={markings, mark=at position 0.55 with {\arrow{stealth reversed}},mark=at position 0.08 with {\arrow{stealth}},mark=at position 0.96 with {\arrow{stealth reversed}}},
        postaction={decorate}] (2.12,-0.5) -- (2.12,-2.5);
        \draw[decoration={markings, mark=at position 0.55 with {\arrow{stealth reversed}},mark=at position 0.08 with {\arrow{stealth reversed}},mark=at position 0.96 with {\arrow{stealth}}},
        postaction={decorate}] (2.48,-2.314) -- (0.76,-1.29);
        \draw[decoration={markings, mark=at position 0.55 with {\arrow{stealth}},mark=at position 0.08 with {\arrow{stealth reversed}},mark=at position 0.96 with {\arrow{stealth reversed}}},
        postaction={decorate}] (2.48,-0.68) -- (0.76,-1.72);
        \draw[fill=black!10] (2.12,-0.9) circle (5pt);
        \draw[fill=black!10] (1.12,-1.5) circle (5pt);
        \draw[fill=black!10] (2.12,-2.1) circle (5pt);
        \node[font=\fontsize{9}{9}] at (2.12,-0.9) {$\textrm{B}$};
        \node[font=\fontsize{9}{9}] at (1.12,-1.5) {$\textrm{B}$};
        \node[font=\fontsize{9}{9}] at (2.12,-2.1) {$\textrm{F}$};
    \end{tikzpicture}
    
    \vskip 0.5cm   
    
    \begin{tikzpicture}[thick, scale=1.8]
        \draw[decoration={markings, mark=at position 0.55 with {\arrow{stealth}},mark=at position 0.08 with {\arrow{stealth}},mark=at position 0.96 with {\arrow{stealth}}},
        postaction={decorate}] (-5,-0.5) -- (-5,-2.5) ;
        \draw[decoration={markings, mark=at position 0.55 with {\arrow{stealth}},mark=at position 0.08 with {\arrow{stealth}},mark=at position 0.96 with {\arrow{stealth}}},
        postaction={decorate}] (-5.36,-2.314) -- (-3.64,-1.29);
        \draw[decoration={markings, mark=at position 0.55 with {\arrow{stealth}},mark=at position 0.08 with {\arrow{stealth}},mark=at position 0.96 with {\arrow{stealth}}},
        postaction={decorate}] (-3.64,-1.72) -- (-5.36,-0.68);
        \draw[fill=black!10] (-5,-0.9) circle (5pt);
        \draw[fill=black!10] (-4,-1.5) circle (5pt);
        \draw[fill=black!10] (-5,-2.1) circle (5pt);
        \node[font=\fontsize{9}{9}] at (-5,-0.9) {$\textrm{F}$};
        \node[font=\fontsize{9}{9}] at (-4,-1.5) {$\textrm{S}$};
        \node[font=\fontsize{9}{9}] at (-5,-2.1) {$\textrm{F}$};
        \node at (-3.4,-1.5) {+};
        \draw[decoration={markings, mark=at position 0.55 with {\arrow{stealth reversed}},mark=at position 0.08 with {\arrow{stealth}},mark=at position 0.96 with {\arrow{stealth}}},
        postaction={decorate}] (-2.8,-0.5) -- (-2.8,-2.5);
        \draw[decoration={markings, mark=at position 0.55 with {\arrow{stealth reversed}},mark=at position 0.08 with {\arrow{stealth}},mark=at position 0.96 with {\arrow{stealth}}},
        postaction={decorate}] (-3.16,-2.314) -- (-1.44,-1.29);
        \draw[decoration={markings, mark=at position 0.55 with {\arrow{stealth reversed}},mark=at position 0.08 with {\arrow{stealth}},mark=at position 0.96 with {\arrow{stealth}}},
        postaction={decorate}] (-1.44,-1.72) -- (-3.16,-0.68);
        \draw[fill=black!10] (-2.8,-0.9) circle (5pt);
        \draw[fill=black!10] (-1.8,-1.5) circle (5pt);
        \draw[fill=black!10] (-2.8,-2.1) circle (5pt);
        \node[font=\fontsize{9}{9}] at (-2.8,-0.9) {$\textrm{B}$};
        \node[font=\fontsize{9}{9}] at (-1.8,-1.5) {$\textrm{B}$};
        \node[font=\fontsize{9}{9}] at (-2.8,-2.1) {$\textrm{B}$};
        \node at (-0.9,-1.5) {=};        
        \draw[decoration={markings, mark=at position 0.55 with {\arrow{stealth}},mark=at position 0.08 with {\arrow{stealth}},mark=at position 0.96 with {\arrow{stealth}}},
        postaction={decorate}] (0.9,-0.5) -- (0.9,-2.5);
        \draw[decoration={markings, mark=at position 0.55 with {\arrow{stealth}},mark=at position 0.08 with {\arrow{stealth}},mark=at position 0.96 with {\arrow{stealth}}},
        postaction={decorate}] (1.26,-2.314) -- (-0.46,-1.29);
        \draw[decoration={markings, mark=at position 0.55 with {\arrow{stealth reversed}},mark=at position 0.08 with {\arrow{stealth reversed}},mark=at position 0.96 with {\arrow{stealth reversed}}},
        postaction={decorate}] (1.26,-0.68) -- (-0.46,-1.72);
        \draw[fill=black!10] (0.9,-0.9) circle (5pt);
        \draw[fill=black!10] (-0.1,-1.5) circle (5pt);
        \draw[fill=black!10] (0.9,-2.1) circle (5pt);
        \node[font=\fontsize{9}{9}] at (0.9,-0.9) {$\textrm{F}$};
        \node[font=\fontsize{9}{9}] at (-0.1,-1.5) {$\textrm{S}$};
        \node[font=\fontsize{9}{9}] at (0.9,-2.1) {$\textrm{F}$};
        \node at (1.4,-1.5) {+};
        \draw[decoration={markings, mark=at position 0.55 with {\arrow{stealth reversed}},mark=at position 0.08 with {\arrow{stealth}},mark=at position 0.96 with {\arrow{stealth}}},
        postaction={decorate}] (3.12,-0.5) -- (3.12,-2.5);
        \draw[decoration={markings, mark=at position 0.55 with {\arrow{stealth reversed}},mark=at position 0.08 with {\arrow{stealth}},mark=at position 0.96 with {\arrow{stealth}}},
        postaction={decorate}] (3.48,-2.314) -- (1.76,-1.29);
        \draw[decoration={markings, mark=at position 0.55 with {\arrow{stealth}},mark=at position 0.08 with {\arrow{stealth reversed}},mark=at position 0.96 with {\arrow{stealth reversed}}},
        postaction={decorate}] (3.48,-0.68) -- (1.76,-1.72);
        \draw[fill=black!10] (3.12,-0.9) circle (5pt);
        \draw[fill=black!10] (2.12,-1.5) circle (5pt);
        \draw[fill=black!10] (3.12,-2.1) circle (5pt);
        \node[font=\fontsize{9}{9}] at (3.12,-0.9) {$\textrm{B}$};
        \node[font=\fontsize{9}{9}] at (2.12,-1.5) {$\textrm{B}$};
        \node[font=\fontsize{9}{9}] at (3.12,-2.1) {$\textrm{B}$};
    \end{tikzpicture}
    \caption{Schematic representation of the three types of Yang-Baxter equations for the case of $U(N)$ scattering. Each line is associated with a given rapidity $\theta_i$, and while arrows oriented upward concern particle states, the ones running downward represent antiparticle states. The $2 \rightarrow 2$ scattering processes are described by the matrices $S$, $F$, $B$ defined in (\ref{eq:S matrix def}), (\ref{eq:F matrix def}) and (\ref{eq:B matrix def}).}
    \label{fig:schematic representation YB eqs}
\end{figure}
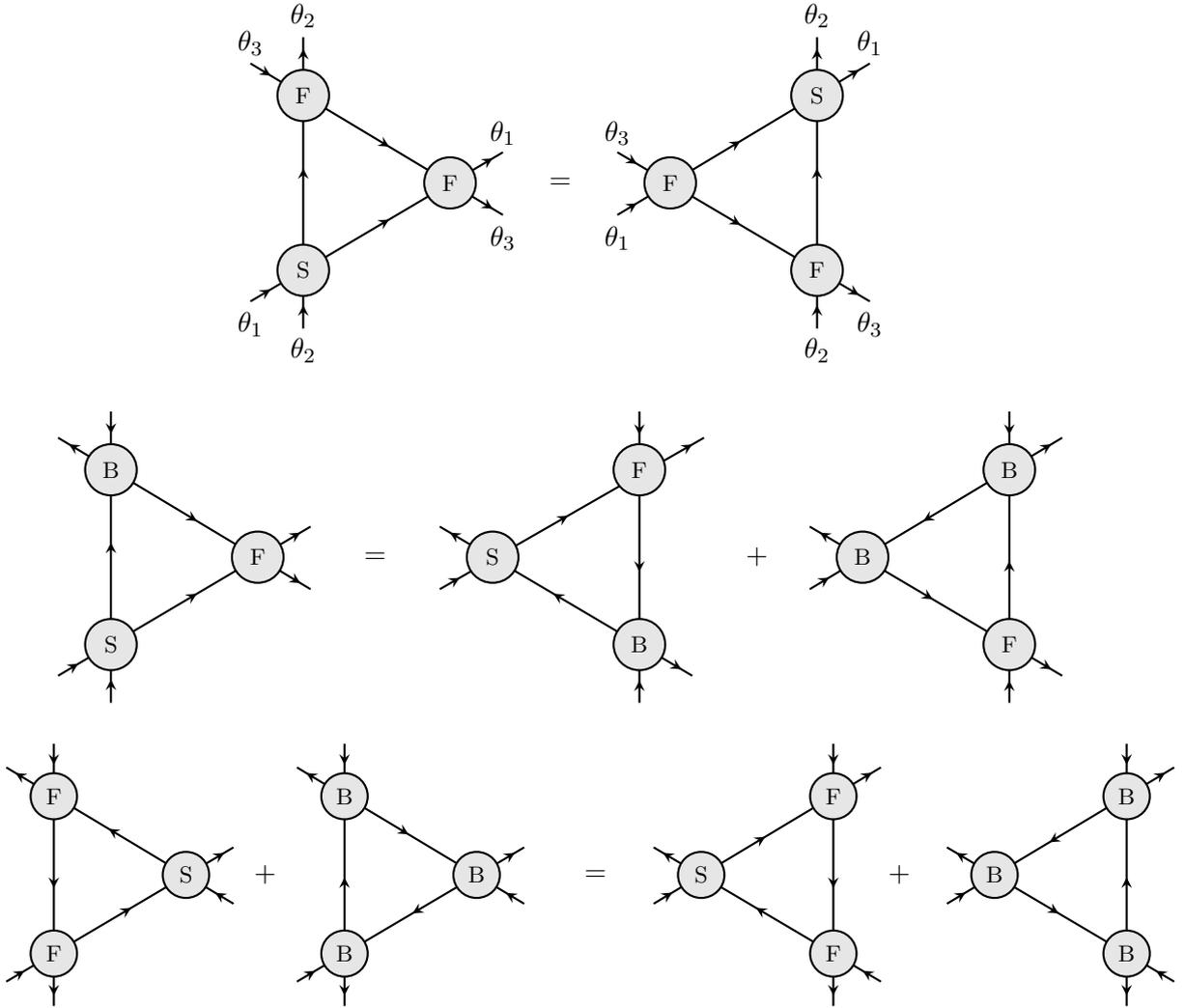

In this Appendix we review the different classes of solutions to the Yang-Baxter equations for particles transforming in the fundamental/antifundamental representation of $U(N)$. The first classification was performed in  \cite{Berg:1977dp}, but some solutions were missed there. Here, we present a complete analysis. As stated in \cite{Berg:1977dp} there are three\footnote{In \cite{Basso:2012bw} the authors write two other kinds of factorization equations stemming from scatterings of the type: $P(\theta_1)\, P (\theta_2) \, P(\theta_3) \longrightarrow P(\theta_3) \, P(\theta_2) \, P(\theta_1)$ and $P(\theta_1)\, P (\theta_2) \, A(\theta_3) \longrightarrow P(\theta_3) \, P(\theta_2) \, A(\theta_1)$. We have checked, however, that the factorization equations coming from these are redundant and as such do not place further constraints which could decrease the number of solutions.} independent types of Yang-Baxter equations for this case, namely
\begin{align}
    &S(\theta_{12}) \, F (\theta_{13}) \, F(\theta_{23}) = F(\theta_{23}) \, F(\theta_{13}) \, S(\theta_{12}) \, , \nonumber \\
    &S(\theta_{12}) \, F(\theta_{13}) \, B(\theta_{23}) = B(\theta_{23}) \, S(\theta_{13}) \, F(\theta_{12})  +  F(\theta_{23}) \, F(\theta_{13}) \, F(\theta_{12}) \, , \label{eq:types of Yang-Baxter eqs for U(N)} \\
    &F(\theta_{12}) \, S(\theta_{12}) \, F(\theta_{23}) + B(\theta_{12}) \, B(\theta_{12}) \, B(\theta_{23}) = F(\theta_{23}) \, S(\theta_{13}) \, F(\theta_{12}) + B(\theta_{23}) \, B(\theta_{13}) \, B(\theta_{12}) , \nonumber
\end{align}
each of which encompasses several independent factorization equations. These can be more easily understood by the schematic representation of figure \ref{fig:schematic representation YB eqs}.

\begin{figure}
    \centering
    \includegraphics[scale=1.4]{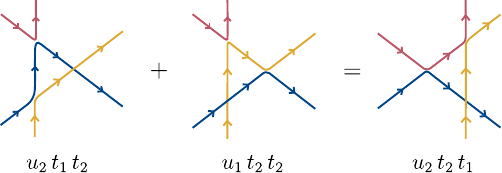}
    \caption{Schematic depiction of the procedure to obtain the Yang-Baxter equations}
    \label{fig:YBeq1 schematic representation}
\end{figure}

A simple way to obtain the Yang-Baxter equations is to label the endpoints of the lines in the schematic representations of figure \ref{fig:schematic representation YB eqs} in pairs and then connect them in all possible ways. In each $2 \rightarrow 2$ scattering process, i.e. when two lines intersect, we identify one of the functions $u_i$, $t_i$, $r_i$ ($i=1,2$) based on the orientation of the arrows. To exemplify this procedure we depict it in a schematic way in figure \ref{fig:YBeq1 schematic representation}. This should then be repeated for different choices of labeling.

In what follows we will present the equations and respective solutions, which resulted from doing this exercise. We will separate the analysis between general $N$ and $N=2$, as the latter contains an extra solution with a free parameter.

\subsection{\texorpdfstring{General $N$}{General N}}
Keeping $N$ general and going through the steps mentioned above we get the following set of equations:
\begin{align}
    &\hspace{4.5cm} u_2 \, t_1 \, t_2 + u_1 \, t_2 \, t_2 = u_2 \, t_2 \, t_1 \, , \label{eq; YB1} \\[5pt]
    &\hspace{4.5cm} u_1 \, t_2 \, r_2 + u_2 \, t_1 \, r_2 = r_1 \, r_2 \, t_1 \, , \label{eq; YB2} \\[5pt]
    &\hspace{3.7cm}u_1 \, t_1 \, r_2 + u_2 \, t_2 \, r_2 =  t_1 \, u_1 \, r_2 + r_1 \, r_2 \, t_2 \, , \label{eq; YB3} \\[5pt]
    &\hspace{4.5cm}t_2 \, u_1 \, r_1 + r_2 \, r_2 \, t_1 = u_1 \, t_2 \, r_1 \, , \label{eq; YB4} \\[5pt]
    &\hspace{4.5cm}t_1 \, u_2 \, r_1 + r_1 \, r_1 \, t_1 = u_2 \, t_1 \, r_1 \, , \label{eq; YB5} \\[5pt]
    &\hspace{5.3cm}u_1 \, t_1 \, r_1  = t_1 \, u_1  \, r_1  \, , \label{eq; YB6} \\[5pt]
    &N \, t_2 \, u_2 \, r_2 + t_1 \, u_2 \, r_2 
    + t_2 \, u_1\, r_2 + t_2\, u_2 \, r_1
    + N \, r_2 \, r_1 \, t_2 + r_1 \, r_1 \, t_2 + r_2 \, r_1 \, t_1 = u_2 \, t_2 \, r_1 \, , \label{eq; YB7} \\[5pt]
    &N \, t_2 \, u_2 \, t_2 + t_2 \, u_2 \, t_1 + t_1 \, u_2 \, t_2 + t_1 \, u_2 \, t_2 + t_2 \, u_1 \, t_2 + N \, r_2 \, r_1 \, r_2 + r_2 \, r_1 \, r_1 + r_1 \, r_1 \, r_2 \nonumber \\
    &+ r_2 \, r_2 \, r_2 = r_1 \, r_2 \, r_1 \label{eq; YB8} \, ,
\end{align}

where in each term the arguments of the functions are $\theta_{12}$, $\theta_{13}$ and $\theta_{23}$, respectively. At this point we would like to introduce the variables $\theta = \theta_{12}$ and $\theta' = \theta_{23}$, such that $\theta + \theta' = \theta_{13}$, as these will be used frequently from now on.

In order to solve the above equations and obtain the solutions of table \ref{table:integrable} we start by drawing our attention to equation (\ref{eq; YB6}). It will only hold true if either
\begin{align}
    r_1 (\theta) = 0 = r_2 (\theta) \, , \label{eq:r1 null condition}
\end{align}
where the second equality comes from crossing, or
\begin{align}
    r_1 (\theta) \neq 0 \neq r_2 (\theta) \quad \text{and} \quad u_1 (\theta) \, t_1 (\theta ') = t_1 (\theta) \, u_1 (\theta ') \, . \label{eq:ratio u1 t1 constant}
\end{align}
As such, below we will divide the solutions by whether they satisfy (\ref{eq:r1 null condition}) or (\ref{eq:ratio u1 t1 constant}), starting with the first.

Before this, let us note that there are zero modes in these equations, so that multiplying a given solution by CDD factors\footnote{These are functions of the form $\text{CDD}_\lambda (\theta)=\frac{\sinh\theta-i\sin\lambda}{\sinh\theta+i\sin\lambda}$ which are crossing symmetric $\text{CDD}_\lambda (i\pi-\theta)=\text{CDD}_\lambda (\theta)$ and saturate unitarity $\text{CDD}_\lambda (-\theta)\text{CDD}_\lambda (\theta)=1$.} gives another integrable amplitude. As usual, we will describe the "minimal solutions" to these equations, given by the set of functions with the minimal amount of zeros and poles. Furthermore, we write the solutions compatible with our no bound states assumption, so that there are no poles in the physical strip.

\subsection*{Class I (Free theory)}
The simplest solution to Yang-Baxter equations is the free propagation case. In this scenario, not only is condition (\ref{eq:r1 null condition}) satisfied as we also have
\begin{align}
    u_2 (\theta) = 0 = t_2 (\theta)\, , \label{eq:null u2 and t2}
\end{align}
such that all equations (\ref{eq; YB1}-\ref{eq; YB8}) hold true. The remaining two components are constant and need to satisfy unitarity and crossing. As such they are given by
\begin{align}
    u_1 (\theta) = 1 = t_1 (\theta) \,\, .
\end{align}

\subsection*{Class II} 
A different solution can be obtained by dropping condition (\ref{eq:null u2 and t2}), implying that $u_2 \neq 0 \neq t_2$, but keeping the assumption (\ref{eq:r1 null condition}) of no backward scattering. From equation (\ref{eq; YB1}) we can then use crossing symmetry, $u_1 (\theta) = t_1 (i \pi - \theta)$, $u_2 (\theta) = t_2 (i \pi - \theta)$, to write everything in terms of $t_1$ and $t_2$. By defining the function
\begin{align}
    p(\theta) = \frac{t_2 (\theta)}{t_1 (\theta)}  \, , \label{eq:ratio t2 t1 function def}
\end{align}
the Yang-Baxter equation (\ref{eq; YB1}) becomes 
\begin{align}
    \frac{1}{p(i \pi - \theta)} - \frac{1}{p(\theta)} + \frac{1}{p(\theta + \theta ')} = 0 \, ,
\end{align}
which admits as solution
\begin{align}
    p(\theta) = \frac{c}{i \pi - \theta} \, , \label{eq:ratio t1 t2 solution}
\end{align}
where $c$ is a constant that remains to be determined. Crossing necessarily implies that
\begin{align}
    \frac{u_2 (\theta)}{u_1 (\theta)} = \frac{c}{\theta} \, . \label{eq:ratio u2 u1}
\end{align}
The remaining factorization equations then fix $c$ to be
\begin{align}
    c = - \frac{2 i \pi}{N} \, . \label{eq:parameter c sol}
\end{align}
Upon replacing this solution in the unitarity equations (\ref{eq:unitarity condition in representation components}) we get one nontrivial equation, namely
\begin{align}
    t_1 (i \pi - \theta) \, t_1 (i \pi + \theta) = \frac{\theta^2}{\theta^2 + \pi^2 \lambda} \, , \label{eq:class II t1 condition}
\end{align}
where $\lambda = \frac{2}{N}$. As described in \cite{Berg:1977dp}, one of the defining equations of the function $f_\lambda(\theta)$, already introduced in (\ref{eq:f_lambda definition}), is
\begin{equation}
    f_\lambda (i \pi - \theta) \, f_\lambda (i \pi + \theta) = \frac{\theta^2}{\theta^2 + \pi^2 \lambda} \, . \label{eq:defining equation of f_lambda}
\end{equation}
Therefore, for this case, we must have
\begin{align}
    t_1(\theta) = f_\lambda (\theta) \, . \label{eq:t1 class II solution}
\end{align}
Classes I and II are the only independent solutions to the Yang-Baxter equations we have found that obey (\ref{eq:r1 null condition}).

\subsection*{Classes III and IV}
We now try to find solutions which satisfy the two conditions in (\ref{eq:ratio u1 t1 constant}). In addition to these, we start by assuming that we also have $u_1 \neq 0 \neq t_1$, such that (\ref{eq:ratio u1 t1 constant}) implies
\begin{align}
     \frac{u_1 (\theta)}{t_1 (\theta)} = k \, , \label{eq:ratio u1 t1 constant 2}
\end{align}
with $k$ being a constant to be determined. The above relation simplifies the factorization equations, in particular (\ref{eq; YB3}), which transforms into
\begin{align}
    u_2 \, t_2 \, r_2 = r_1 \, r_2 \, t_2 \, . \label{eq:classes III and IV simplified YB eq}
\end{align}
One possible solution for this equation is
\begin{align}
    u_2 (\theta) = r_1 (\theta) \quad , \quad t_2 (\theta) = r_2 (\theta) \, . \label{eq:u2er1 and t2er2}
\end{align}
Thus, we replace this in the Yang-Baxter equations, we use crossing to write everything in terms of $r_1$ and $t_1$ and similarly to what was done for class II we define a ratio function
\begin{align}
    \rho(\theta) = \frac{r_1(\theta)}{t_1 (\theta)} \, . \label{eq:ratio r1 t1 function def}
\end{align}
After doing this, equation (\ref{eq; YB5}) becomes
\begin{align}
    \frac{1}{\rho (\theta)} + \frac{1}{\rho (\theta ')} = \frac{1}{\rho(\theta + \theta')} \, , \label{eq:defining eq for ratio of r1 t1 cIII and IV}
\end{align}
whose solution is given by
\begin{align}
    \rho (\theta) = \frac{b}{\theta} \, , \label{eq:rho solution}
\end{align}
where $b$ is an arbitrary constant. Combining crossing together with the relation (\ref{eq:ratio u1 t1 constant 2}) we also get that
\begin{align}
    r_2 (\theta) = \frac{b \cdot k}{i \pi - \theta} \, t_1 (\theta) \, .
\end{align}
Once we replace all of the relations obtained so far in the Yang-Baxter equations we get two remaining independent equations coming from (\ref{eq; YB1}) (and the same equation from (\ref{eq; YB2})) and (\ref{eq; YB7}) (and a similar equation from (\ref{eq; YB8})). The first of these tells us that the following condition needs to be satisfied
\begin{align}
    k \, (k^2-1) = 0 \, . \label{eq:condition on k for classes III and IV}
\end{align}
Given that the solution $k=0$ is inconsistent with crossing and our assumption that $r_2 \neq 0$, there are only two possible scenarios:
\begin{align}
    k=1 \quad \text{or} \quad k=-1 \,\, . \label{eq:solutions to k} 
\end{align}
\subsubsection*{Class III ($k=1$)}
If we choose $k=1$ the non-trivial equation coming from (\ref{eq; YB7}) fixes the constant $b$ to be
\begin{align}
    b = - \frac{i \, \pi}{N-1} \, .
\end{align}
The only thing missing is the function $t_1 (\theta)$, in terms of which everything else is written. To obtain this we use the unitarity conditions (\ref{eq:unitarity condition in representation components}) expressed in terms of the components $u_i$, $t_i$, $r_i$ ($i=1,2$) and use all the results found so far. If we do this, we get
\begin{align}
    t_1 (\theta) \, t_1(- \theta) = \frac{\theta^2}{\theta^2 + \pi^2 \, \lambda^2} \, , \label{eq:defining eq t1 cIII}
\end{align}
where $\lambda = 1/(N-1)$. Additionally, we get another constraint for $t_1(\theta)$ if we use both crossing between $t_1$ and $u_1$ and the fact that they are related by a constant in this case. In particular, we arrive at the condition
\begin{align}
    t_1(i \pi - \theta) = t_1 (\theta) \, . \label{eq:auto-crossing of t1}
\end{align}
The constraints (\ref{eq:defining eq t1 cIII}) and (\ref{eq:auto-crossing of t1}) are simultaneously satisfied if $t_1$ is given by
\begin{align}
    t_1(\theta) = f_\lambda (\theta) \, f_{\lambda}(i \pi - \theta) \, . \label{eq:t1 sol class III}
\end{align}
This results from (\ref{eq:defining equation of f_lambda}) and another of the defining equations of $f_\lambda (\theta)$ \cite{Berg:1977dp}, namely
\begin{align}
    f_\lambda (\theta) \,  f_\lambda (-\theta) = 1 \, . \label{eq:function f "unitarity" condition}
\end{align}

\subsubsection*{Class IV ($k=-1$)}
If on the other hand we choose $k=-1$, (\ref{eq; YB7}) forces $b$ to be
\begin{align}
    b = \frac{i \pi}{N + 1} \, .
\end{align}
Analogously to the previous case, the unitarity conditions then tell us that the function $t_1$ must obey
\begin{align}
    t_1 (\theta) \, t_1(-\theta) = \frac{\theta^2}{\theta^2 + \pi \lambda^2} \, , \label{eq:condition t1 class IV}
\end{align}
with $\lambda = 1/(N+1)$. In addition, crossing and the relation between $t_1$ and $u_1$ imposes that
\begin{align}
    t_1 (i \pi - \theta) = - t_1(\theta) \, . \label{eq:anti-crossing t1 class IV}
\end{align}
It is possible to see that
\begin{align}
    t_1 (\theta) = f_\lambda (\theta) \, f_{\lambda} (i \pi - \theta) \, i \, \tanh\left( \frac{\theta - i  \frac{\pi}{2}}{2}\right) \, , \label{eq:t1 sol class IV}
\end{align}
satisfies both constraints (\ref{eq:condition t1 class IV}, \ref{eq:anti-crossing t1 class IV}), with the factor $i \, \tanh\left[\left(\theta - i  \frac{\pi}{2}\right)/2\right]$ in particular taking care of the anti-crossing property (\ref{eq:anti-crossing t1 class IV}). We point out, however, that it also places a zero\footnote{The solution in \cite{Berg:1977dp} inserts a pole at $\theta = i \pi/2$. Given that we are not assuming any bound states in the spectrum of the theory, we cannot have such poles. To go from the expression found in \cite{Zamolodchikov:1978xm} to the one found here, we multiply by a CDD factor $\text{CDD}_{\pi/2}(\theta)=\frac{\tanh\frac{1}{2}(\theta-i\pi/2)}{\tanh\frac{1}{2}(\theta+i\pi/2)}$. } at the crossing-symmetric point $\theta = i \, \pi/2$.

\subsection*{Class V}
Another possibility for (\ref{eq:ratio u1 t1 constant}) to hold is by having\footnote{If one of them is null the other also needs to be by crossing.}.
\begin{align}
    u_1 = 0 = t_1 \, .
\end{align}
Setting this in the Yang-Baxter equations we once again conclude that (\ref{eq:u2er1 and t2er2}) needs to be satisfied. After using (\ref{eq:ratio u1 t1 constant}) and (\ref{eq:u2er1 and t2er2}) we get one nontrivial equation for the ratio of functions
\begin{align}
    a(\theta)  = \frac{r_2(\theta)}{r_1 (\theta)} \, , \label{eq:ratio r2 r1 def}
\end{align}
coming from (\ref{eq; YB7}) and (\ref{eq; YB8}), namely
\begin{align}
    a(\theta) - a(\theta + \theta') + a(\theta') + 2\, N \, a(\theta) \, a(\theta ') + a(\theta) \, a(\theta + \theta') \,  a(\theta ') = 0 \, . \label{eq:ratio r2 r1 defining equation}
\end{align}
Due to crossing between $r_1$ and $r_2$, $a(\theta)$ must satisfy $a(i \pi - \theta) = 1/a(\theta)$. It can be shown that the general solution to this equation is
\begin{align}
    a(\theta) = \frac{\sin \mu \, \theta}{\sin \mu \, (i \pi - \theta)} \, , \label{eq:ratio r2 r1 solution}
\end{align}
where $\mu$ is fixed by (\ref{eq:ratio r2 r1 defining equation}) to be
\begin{align}
    \cosh(\pi \mu) = N \, . \label{eq:mu def class V}
\end{align}
We point out that our solution for $a(\theta)$ in (\ref{eq:ratio r2 r1 solution}) is the inverse of the one written for class V in \cite{Berg:1977dp}. However, the former is the correct one since poles on the physical strip (outside the imaginary axis for the rapidity)  would lead to a causality violation. Moreover, their expression does not satisfy Yang-Baxter equations. 

At this point only $r_1$ remains to be fixed. In order to do this we use the only independent unitarity condition for this case
\begin{align}
    r_1 (\theta) \, r_1 (-\theta) = 1 \, , \label{eq:unitarity class V}
\end{align}
which together with (\ref{eq:ratio r2 r1 solution}) and crossing leads to
\begin{align}
    r_1 (-i \pi  + \theta) \, r_1 (-i \pi - \theta) = - \frac{\sin \mu(i \pi - \theta) \, \sin \mu (i \pi + \theta)}{\sin ^2 \, \mu \, \theta} \, . \label{eq:defining eq r1 class V}
\end{align}
Given that the sine function can be expressed as
\begin{align}
    \sin x = \prod_{n=1}^{\infty} \left( 1- \frac{x^2}{\pi^2 \, n^2} \right) \, , \label{eq:product definition of sine}
\end{align}
one can write (\ref{eq:defining eq r1 class V}) in such a way that it is relatively simple to make a connection with (\ref{eq:defining equation of f_lambda}) and show that the solution is given by
\begin{align}
    r_1 (\theta) = \prod_{k=-\infty}^{\infty} \frac{f_{1-i\, k/\mu}(\theta)}{f_{i \, k/\mu} (\theta)} \, . \label{eq:class V r1 solution} 
\end{align}

\subsection*{Class VI}
To get a different solution we recognize that (\ref{eq:classes III and IV simplified YB eq}) is also satisfied if we have
\begin{align}
    u_2 (\theta) = c(\theta) \, r_1 (\theta) \quad , \quad t_2 (\theta) = c(i \pi - \theta) \, r_2 (\theta) \, , \label{eq:relation between components with extra function}
\end{align}
where $c(\theta)$ is an arbitrary function to be determined. The second equation comes from crossing.
Using these relations, equation (\ref{eq; YB3}) requires that $c(\theta)$ obeys
\begin{align}
    c(\theta) \, c(\theta' - \theta) = c(\theta') \, , \label{eq;defining eq c(theta)}
\end{align}
which is satisfied by $c(\theta) = e^{i \mu \theta}$, where $\mu$ is a (in principle complex) parameter to be determined. However, for this class unitarity demands that $\| u_2(\theta) \|^2 = 1$ and $\| r_1(\theta) \|^2 = 1 $, which in turn implies that $\| c(\theta)^2 \|=1$, so that $\mu\in\mathbb R$. Similarly to the previous class, $r_1$ is given by (\ref{eq:class V r1 solution}) and $r_2$ is related to $r_1$ by (\ref{eq:ratio r2 r1 def}, \ref{eq:ratio r2 r1 solution}). Using all of these relations as input for the Yang-Baxter equations, we arrive at last at
\begin{align}
    e^{\pi \, \mu} = N \, , \label{eq:mu solution class VI}
\end{align}
stemming from (\ref{eq; YB7}) and (\ref{eq; YB8}).

\subsection*{Class VII}
Finally, there is another solution not considered in \cite{Berg:1977dp} where $u_1 = 0 = t_1$ and $u_2 = 0 = t_2$. Setting $r_2$ related to $r_1$ as in class V by (\ref{eq:ratio r2 r1 solution}), we get a similar equation that involves $r_1$, the only difference being $N \rightarrow N/2$. Consequently, in this case the constant $\mu$ gets fixed by (\ref{eq; YB8}) to be
\begin{align}
    \cosh{\pi \mu} = \frac{N}{2} \, . \label{eq:mu def for class VII}
\end{align}
Given that the unitarity condition remains unchanged from class V, the function $r_1 (\theta)$ has the same form as in (\ref{eq:class V r1 solution}). 

\subsection{\texorpdfstring{$N$=2}{N=2}}
We now turn our attention to the case of $N=2$. Naturally, all the solutions obtained before are still valid for $N=2$ but as it turns out this case admits an extra solution, more specifically a family of solutions, characterized by a continuous parameter, which was first derived in \cite{Wiegmann:1985jt} (see also \cite{Basso:2012bw}).

Repeating the exercise mentioned before to arrive at the Yang-Baxter equations, we obtain the set of independent equations\footnote{We point out that (\ref{eq:N=2 eq 1}) and (\ref{eq:N=2 eq 4}) do not come from the type of Yang-Baxter equations in figure \ref{fig:schematic representation YB eqs} but rather from the ones we mentioned in the footnote below (\ref{eq:types of Yang-Baxter eqs for U(N)}). In particular, (\ref{eq:N=2 eq 1}) is of the type $S \, S \, S = S \, S \, S$ while (\ref{eq:N=2 eq 4}) corresponds to $S \, B  \, S = F \, B \, F + B \, S \, B$, where the order of arguments is $\theta_{12}$, $\theta_{13}$, $\theta_{23}$ respectively. We have chosen these equations to match with previous literature  \cite{Basso:2012bw} and because in this $N=2$ case they turn out to be more straightforward to solve.}

\begin{align}
    & u_1 \, u_2 \, u_2 + u_2 \, u_2 \, u_1 = u_2 \, u_1 \, u_2 \, , \label{eq:N=2 eq 1} \\
    & u_1 \, t_2 \, t_2 + u_2 \, t_2 \, t_2 = u_2 \, t_2 \, t_2 \, . \label{eq:N=2 eq 2} \\
    &u_1 \, t_1 \, r_1 + u_2 \, t_1 \, r_1 = t_1 \, u_1 \, r_1 + t_1 \, u_2 \, r_1 + r_1 \, r_1 \, t_1 \,  , \label{eq:N=2 eq 3}\\
    &u_1 \, r_1 \, u_1 + u_1 \, r_1 \, u_2 + u_2 \, r_1 \, u_1 + u_2 \, r_1 \, u_2 = t_1 \, r_1 \, t_1 + r_1 \, u_1 \, r_1 + r_1 \, u_2 \, r_1 \, , \label{eq:N=2 eq 4}\\
    &u_1 \, t_1 \, r_2 + u_2 \, t_1 \, r_2 + u_1 \, t_2 \, r_2 + u_2 \, t_2 \, r_2  = t_1 \, u_1 \, r_2 + r_1 \, r_2 \, t_1 + r_1 \, r_2 \, t_2 \, , \label{eq:N=2 eq 5} \, .
\end{align}
Starting with the first two equations (\ref{eq:N=2 eq 1}) and (\ref{eq:N=2 eq 2}), and taking into account the crossing relations between $u_i$ and $t_i$ $(i=1,2)$, it is simple to see that these admit as solutions
\begin{equation}
    u_2(\theta)  = \frac{c}{\theta} \, u_1 (\theta) \quad , \quad t_2(\theta)  = \frac{c}{i \pi - \theta} \, t_1 (\theta) \, , \label{eq:u1, u2 relation and t1, t2 relation}
\end{equation}
where $c$ is an arbitrary constant.

We now shift our focus to equations (\ref{eq:N=2 eq 3}) and (\ref{eq:N=2 eq 4}) and re-express them as
\begin{align}
    & m (\theta + \theta ') + n(\theta) \, n(\theta') = m (\theta) \, m(\theta') \, \label{eq:m and n eq 1} \\
    & n(\theta) \, m (\theta + \theta ') + n(\theta ') = m(\theta) \, n(\theta + \theta ') \, , \label{eq:m and n eq 2}
\end{align}
where we have made the following definitions
\begin{align}
    m(\theta) = \frac{u_1(\theta) + u_2(\theta)}{r_1(\theta)} \quad , \quad n(\theta) = \frac{t_1(\theta)}{r_1(\theta)} \, . \label{eq:m and n definitions}
\end{align}
To give an idea on how to solve (\ref{eq:m and n eq 1}) and (\ref{eq:m and n eq 2}) we follow the logic of appendix A in \cite{Zamolodchikov:1978xm}. Thus, we start by setting either $\theta = 0$ or $\theta ' = 0$ from which we find that we must have: $n(0) = 0$, $m(0) = 1$. Next, we differentiate with respect to $\theta '$, for example, and set $\theta ' = 0$ afterwards, thus obtaining
\begin{align}
    & m' (\theta) + \beta \, n (\theta) = \alpha \, m (\theta) \, \, \label{eq:m and n diff eq 1}\\
    & m(\theta) \, n' (\theta) =  n(\theta) \, m'(\theta) + \beta \, , \label{eq:m and n diff eq 2}
\end{align}
where the following definitions have been made
\begin{equation}
    \alpha = m'(0) \quad , \quad \beta = n'(\theta) \, . \label{eq:alpha and beta definitions}
\end{equation}
These equations can be transformed into
\begin{align}
    & m''(\theta) \, m(\theta) - m'(\theta)^2 + \beta^2 = 0 \, , \label{eq:second order diff eq m} \\
    & n(\theta) = \frac{\alpha \, m(\theta) - m' (\theta)}{\beta} \, . \label{eq:n in terms of m}
\end{align}
In order to solve these we must remember to impose the appropriate boundary conditions: $m(0) = 1$, $m'(0)=\alpha$, $n(0) = 0$, $n'(0)=\beta$. It is possible to see that the solutions for $m(\theta)$ and $n(\theta)$ satisfying both (\ref{eq:second order diff eq m}) and (\ref{eq:n in terms of m}), and the aforementioned boundary conditions are given by
\begin{equation}
    \begin{aligned}
        &m(\theta) = - i \, \frac{\sinh\left( \frac{i \pi - \theta}{p} \right)}{\sin \left( \frac{\pi}{p} \right)} \quad , \quad n(\theta) = - i \, \frac{\sinh\left( \frac{ \theta}{p} \right)}{\sin \left( \frac{\pi}{p} \right)} \, ,
    \end{aligned}
    \label{eq:m and n solutions}
\end{equation}
where we have made the following choice of re-parametrization of $\alpha$ and $\beta$:
\begin{equation}
    \alpha = i \, \frac{\cot{\frac{\pi}{p}}}{p} \quad , \quad \beta = - i \, \frac{\csc{\frac{\pi}{p}}}{p} \, ,
\end{equation}
in order to simplify the expressions. Indeed, with the above rewriting of these parameters they satisfy the identity
\begin{equation}
    \alpha^2 - \beta^2 = \frac{1}{p^2} \, .
\end{equation}


By making use of what we have found so far, in particular (\ref{eq:u1, u2 relation and t1, t2 relation}) and (\ref{eq:m and n solutions}), the remaining Yang-Baxter equation (\ref{eq:N=2 eq 5}) fixes the parameter $c$ to be
\begin{equation}
    c = - i \, \pi \, .
\end{equation}
The last missing piece is then $u_1 (\theta)$. Since all Yang-Baxter equations are solved at this point, we turn to the unitarity condition (\ref{eq:unitarity condition in representation components}), which imposes
\begin{equation}
    u_1 (\theta) \, u_1 (- \theta) = \frac{\theta^2}{\pi^2 + \theta^2} \, . \label{eq:u1 condition for N=2 continuous solution}
\end{equation}
Interestingly, this is exactly (\ref{eq:defining equation of f_lambda}) for $\lambda=1$, i.e. one of the defining equations of $f_{\lambda=1} (\theta)$. However, we also need to satisfy crossing between $u_1(\theta)$ and $t_1(\theta)$, which in this case looks like
\begin{equation}
    t_1(\theta) = - \frac{\sinh \left( \frac{\theta}{p} \right)}{\sinh \left( \frac{i \pi - \theta}{p} \right)} \, \left( 1 + \frac{\pi}{i \theta} \right) \, u_1(\theta) = u_1 (i \pi - \theta) \, . \label{eq:crossing t1, u1 for N=2}
\end{equation}
Bearing in mind that $f_1 (\theta)$ obeys $f_1 (\theta)= - \left( 1 + \frac{\pi}{i \theta} \right)f_1 (i \pi - \theta)$ ,
one can write the solution to (\ref{eq:u1 condition for N=2 continuous solution}) as
\begin{equation}
    u_1 (\theta) = \frac{\theta}{\theta - i \pi} \, f_1(\theta) \, \xi_p (\theta) \, , \label{eq:u1 solution N=2}
\end{equation}
where $\xi_p(\theta)$ necessarily needs to satisfy
\begin{align}
    & \xi_p (\theta) \, \xi_p(-\theta) = 1 \, , \label{eq:xi condition 1}\\
    & \xi_p (i \pi - \theta) = \frac{\sinh \left( \frac{\theta}{p} \right)}{\sinh \left( \frac{i \pi - \theta}{p} \right)} \, \xi_p(\theta) \, . \label{eq:xi condition 2}
\end{align}
The solution to these equations is well known \cite{Zamolodchikov:1978xm}  and can be expressed in an integral representation as  (see e.g. section 18.8 of \cite{Mussardo:2010mgq})

\begin{equation}
    \xi_p(\theta) = - \exp\left[ i \, \int_{0}^{\infty} \frac{d\omega}{\omega}  \frac{\sinh \left( \frac{\pi \,
    \omega \, (p - 1)}{2} \right)}{\sinh\left( \frac{\pi \, \omega \, p}{2} \right) \, \cosh \left( \frac{\pi \, \omega}{2} \right)}\right] \, . \label{eq:xi_p definition}
\end{equation}
This is the Sine-Gordon soliton scattering phase as mentioned in \cite{Basso:2012bw}.

The $N=2$ continuous parameter solution is thus defined by
\begin{equation}
    \begin{aligned}
        &u_1 (\theta) = \frac{\theta}{\theta - i \, \pi} \, f_1(\theta) \, \xi_p(\theta) = t_1 (i \pi - \theta) \, , \\
        &u_2 (\theta) = \frac{\pi}{i \, \theta} \, u_1(\theta) = t_2 (i \pi - \theta) \, , \\
        &r_1 (\theta) = i \, \frac{\sin\left( \frac{\pi}{p}\right)}{\sinh\left( \frac{i \, \pi - \theta}{p}\right)} \, \left( 1 + \frac{\pi}{i \, \theta} \right)  \, u_1(\theta) = r_2 (i \pi - \theta)\, .
    \end{aligned}
    \label{eq:u,t,r expressions continuous parameter solution}
\end{equation}
It is interesting to note that this family of solutions parametrizes a continuous line on the surface of the $U(2)$ monolith which interpolates between class II ($p \rightarrow 1$) and class III ($p \rightarrow \infty$). Consequently, due to the $S \rightarrow - S$ symmetry we have a similar line of solutions on the other side of the boundary. However, for $p<1$ the solution has poles inside the physical strip associated with bound states and as such these amplitudes are found outside the monolith. 


\section{Integrable form factors and c-sum rule}\label{app:form factor expressions}
Here we briefly review the relation between amplitudes, form factors and central charge used in section~\ref{sec:Discussion}, for more details see e.g. \cite{Cordova:2023wjp}. The starting point is Zamolodchikov's c-theorem \cite{Zamolodchikov:1986gt} which can be expressed as the following sum rule for the trace of the stress tensor $\Theta$
\begin{equation}
    c^\text{UV}-c^\text{IR}=12\pi \int\limits_0^\infty ds \, \frac{\rho_\Theta(s)}{s^2}\,,
\end{equation}
where $\rho_\Theta$ is the spectral density defined as $2\pi\,\rho_\Theta(s)=\int d^2x\,e^{-ip\cdot x}\langle\Theta(x)\Theta(0)\rangle$. Since we are dealing with gapped theories, we have $c^\text{IR}=0$ and the sum rule computes directly the UV central charge. By inserting a complete set of states in $\langle\Theta(x)\Theta(0)\rangle$, we can read off the $n$-particle contribution to the central charge $c=\sum_n c_n$.\footnote{In absence of bound states the first term in this sum is the two-particle contribution $c_2$.} We will be interested in the two-particle contribution to this sum rule $c_2$, since we can compute it from the two-particle amplitudes at hand and associated form factors. The sum rule then becomes
\begin{equation}
    c=c_2+\ldots =3\int_{4m^2}^\infty ds\, \frac{|F_\Theta(s)|^2}{s\sqrt s\sqrt{s-4m^2}} + \ldots\,,
\end{equation}
where $F_\Theta(s)$ is the two-particle form factor.

If the theory is integrable, we can straightforwardly compute the two-particle form factor using Watson's equation $F(s)=S(s) F^*(s)$ \cite{Watson:1954uc}.\footnote{It is also possible to bootstrap the $n$-particle form factors in these theories \cite{Smirnov:1992vz}, although their computation is more involved.} A simple recipe follows from writing the corresponding amplitude with an integral representation of the form (recall $\theta$ is the rapidity variable satisfying $s=4m^2\cosh^2\theta/2$)
\begin{equation}
    S(\theta) = \exp\left( \int_{0}^{\infty} \frac{dt}{t} \mathbb f(t) \sinh \frac{t \theta}{i \pi} \right) \, ,
    \label{eq:integral rep of amplitude}
\end{equation}
where the function $\mathbb f(t)$ can be read off from the zeros and poles of the amplitude as explained in appendix A of \cite{Cordova:2023wjp}. Then a minimal solution to Watson's equation is given by\footnote{The full form factor can be expressed as $F(\theta) = \frac{Q(\theta)}{D(\theta)} \, F_{\text{min}}(\theta)$, where $Q$ and $D$ are polynomials in $\cosh \theta$ (see e.g. \cite{Mussardo:2010mgq} for more details). The first one is operator dependent and the latter is fixed by the S-matrix poles. In the examples discussed below $F_{\text{min}}(\theta)$ with a suitable normalization is enough to describe $F_\Theta(\theta)$.}
\begin{equation}
    F_{\text{min}}(\theta) = \mathcal{N} \exp\left( \int_{0}^{\infty} \frac{dt}{t} \frac{\mathbb f(t)}{\sinh t} \, \sin^2\left[ \frac{t \, (i \pi- \theta)}{2 \pi}  \right]\right) \, .
    \label{eq:F min in terms of f(t)}
\end{equation}
where $\mathcal{N}$ is the normalization.

Going back to our $U(N)$ amplitudes, we now want to use the ingredients above to compute the central charge sum rule up to different energies as in \eqref{eq:two particle contribution to central charge}. The amplitudes we study here are the periodic solutions given by classes V-VII in table~\ref{table:integrable}. In particular, we are interested in the sing+ channel, as the trace of the stress tensor is a singlet of the $U(N)$ symmetry with even parity. These are given by
\begin{align}
    &S^{\text{V}}_\text{sing+}(\theta) = s(\mu_{\text{V}}|\theta)\, , \quad &\mu_{\text{V}} =& \frac{\text{arccosh} \, N}{\pi} \label{eq:singlet plus V} \, , \\
    &S^{\text{VI}}_\text{sing+}(\theta) = e^{- i \mu_{\text{VI}}\, \theta} \, s(\mu_{\text{VI}}| \theta)\, , \quad &\mu_{\text{VI}} =& \frac{\log N}{\pi} \label{eq:singlet plus VI} \, ,\\
    &S^{\text{VII}}_\text{sing+}(\theta) = s(\mu_{\text{VII}}|\theta) \,, \quad &\mu_{\text{VII}} =& \frac{\text{arccosh} \, N/2}{\pi} \label{eq:singlet plus VII} \, ,
\end{align}
where we have defined
\begin{align}
    &s(\mu|\theta) = \frac{\sinh \mu \left( \pi - i \, \theta \right)}{\sinh \mu \left( \pi + i \, \theta \right)} \, \prod_{k \in \mathbb{Z}} F_{\pi + \frac{i k \pi}{\mu}} (-\theta) \,  \label{eq:small s function definition} \, ,
\end{align}
and $F_{\lambda}(\theta)$ is given by a ratio of the building block functions $f_a(\theta)$ in (\ref{eq:f_lambda definition})
\begin{align}
    F_\lambda (\theta) = \frac{f_{\frac{\lambda}{\pi}}(-\theta)}{f_{\frac{\lambda}{\pi}-1}(-\theta)} = \frac{\Gamma\left(\frac{\lambda + i \theta}{2\pi}\right) \, \Gamma\left(\frac{\lambda - i \theta + \pi}{2\pi}\right)}{\Gamma\left(\frac{\lambda - i \theta}{2\pi}\right) \, \Gamma\left(\frac{\lambda + i \theta + \pi}{2\pi}\right)} \, . \label{eq:F building block}
\end{align}
The expression formally coincides with the singlet channel amplitude of the $O(N)$ periodic Yang-Baxter solution and the associated $\mathbb f(t)$ and $F_\text{min}(\theta)$ were written down in \cite{Cordova:2023wjp}. Using their results, we have:
\begin{equation}
    \mathbb f^\diamond _\text{sing+}(t)=\mathbb f(\mu_\diamond|t)\,,\quad \text{with 
}\; \diamond=\text{V, VI, VII}\,.
\end{equation}
where we have defined the kernel associated to \eqref{eq:small s function definition}
\begin{equation}
    \mathbb f(\mu|t) = \frac{4 \, \pi \, \mu \, e^{-t}}{1 + e^t} \sum_{p=0}^{\infty} \frac{\delta(t - 2 \pi \mu p)}{1 + \delta_{p,0}} \, .
    \label{eq:f(t) for singlet plus}
\end{equation}
Similarly, the minimum form factors are expressed in terms of 
\begin{equation}
     F_{\text{min}}(\mu|\theta) = 
     \mathcal{N}   \times 
     \exp \left[ - \frac{(\pi + i \theta)^2 \, \mu}{4 \pi} + \sum_{p=1}^{\infty} \frac{2 e^{-2 \pi \mu p} \, \sin^2\left[(i \pi - \theta) \mu p \right]}{\sinh(2 \pi \mu p) \left(1 + e^{2 \pi \mu p}\right) \, p}\right] \, ,
    \label{eq:F min for singlet plus}
\end{equation}
with the corresponding $\mu$ parameters. The two-particle form factors for $\Theta$ then read
\begin{equation}
     F^\text{V}_\Theta(\theta)= F_{\text{min}}(\mu_\text{V}|\theta)\,,\quad
    F^\text{VI}_\Theta(\theta)= e^{i\mu_\text{VI}\,\theta/2} F_{\text{min}}(\mu_\text{VI}|\theta)\,,\quad
    F^\text{VII}_\Theta(\theta)= F_{\text{min}}(\mu_\text{VII}|\theta)\,,
\end{equation}
with the normalization $\mathcal N=-2\sqrt{2N}m^2$, which can be computed as in \cite{Correia:2022dyp,Cordova:2023wjp}. Note that the phase in $F^\text{VI}_\Theta(\theta)$ is irrelevant for the central charge sum rule. 

To generate the plots in figure~\ref{fig:walking}, we need to consider the amplitudes in table~\ref{table:integrable} with an overall minus sign. For the form factor this implies we should multiply by an overall factor $-i\sinh\tfrac{\theta}{2}$
\begin{equation}
      F^\diamond_\Theta(\theta) \underset{S^\diamond\rightarrow-S^\diamond} {\longrightarrow} -i\sinh\tfrac{\theta}{2}\, F^\diamond_\Theta(\theta)\,, \quad\text{with }\; \diamond=\text{V,VI,VII}\, . 
\end{equation}

Classes V and VI are different in general but coincide for $N=1$ as $\mu_\text{V},\mu_\text{VI}\rightarrow0$. This means that the walking behavior of the central charge in class VI is qualitative the same as class V in figure~\ref{fig:walkingV} for $N\gtrsim1$.

\section{Primal and dual codes}\label{app:codes}
We leave here the \texttt{Mathematica} codes for both the primal and dual approaches that were used to generate all the plots in this work. In particular, we write down the parts which do the optimization and return a point on the boundary of the allowed space. To generate the full figures we simply have to run these in different directions and plot the points at the end.

The first step is to set the value of $N$. Seeing as here we have studied both $O(2)$ and $O(7)$ we either run
\small\begin{verbatim}
n = 2 or n = 7;
\end{verbatim}\normalsize
accordingly. We also define the rho variables used in both ans{\"a}tze (\ref{eq:rho ansatz for the S-matrix}) and (\ref{eq:rho ansatz for K}),
\small\begin{verbatim}
r[s0_,s_]:=(Sqrt[4-s0]-Sqrt[4-s])/(Sqrt[4-s0]+Sqrt[4-s]);
\end{verbatim}\normalsize
and we set the crossing matrix,
\small\begin{verbatim}
c={{1/(2n),(-1+n^2)/(2n),-(1/(2 n)),-((-1+n^2)/(2 n)),(1+n)/2,(1-n)/2},
{1/(2n),-(1/(2n)),-(1/(2n)),1/(2n),1/2,1/2},
{-(1/(2n)),-((-1+n^2)/(2n)),1/(2n),(-1+n^2)/(2n),(1+n)/2,(1-n)/2},
{-(1/(2 n)),1/(2n),1/(2n),-(1/(2n)),1/2,1/2},
{1/(2n),(-1+n)/(2n),1/(2n),(-1+n)/(2n),0,0},
{-(1/(2n)),(1+n)/(2n),-(1/(2n)),(1+n)/(2n),0,0}};
\end{verbatim}\normalsize

After this is done, we can move on to the primal and dual implementations. We shall start with the former.

\subsection{Primal}
Although we have implemented both the normal and radial approaches (see \cite{Cordova:2019lot}) to the primal method, we will only write the latter since it was the one we used the most. Throughout all the functions we will define, we shall leave the number of coefficients of the ansatz \textit{nmax} and the size of the grid of points where we evaluate unitarity \textit{ngrid} as arguments. This way we may vary them freely in order to obtain the best results in a reasonable amount of time.

We start by constructing the grid of points for $s \ge 4m^2$ where we will impose unitarity
\small\begin{verbatim}
chebyshevGrid[n_][a_,b_]:=(a+b)/2-(b-a)/2*Table[Cos[(2j-1)/(2n)\[Pi]],{j,1,n}];
sgrid[NN_]:=sgrid[NN]=N[Table[8/(1+Cos[\[Phi]]),{\[Phi],
N[chebyshevGrid[NN][0,\[Pi]]]}],50];
\end{verbatim}\normalsize
Then, we proceed with setting the ansatz for the S-matrix (\ref{eq:rho ansatz for the S-matrix})
\small\begin{verbatim}
Sn[s_,nmax_]:={singP[0],(-singP[0]+n^2*sym[0]+n(-singM[0]+sym[0]))/(-1+n^2), 
singM[0],-((singM[0]+n(singP[0]-(1+n)sym[0]))/(-1+n^2)),sym[0],
-((singM[0]+singP[0]-(1 + n)sym[0])/(-1+n))}+
Sum[{singP[n], adjP[n], singM[n], adjM[n], sym[n], anti[n]}*
r[2,s]^n+c.{singP[n],adjP[n],singM[n],adjM[n],sym[n],anti[n]}*r[2,4-s]^n,
{n, 1, nmax}];
\end{verbatim}\normalsize
where it can be observed that we already used crossing to relate some of the components of the constant part. Additionally, we specify the free parameters which will be fixed by maximization of the functional,
\small\begin{verbatim}
varsn[nmax_] := Variables[Sn[3.5, nmax]];
\end{verbatim}\normalsize
Next, we evaluate our ansatz in the previously defined grid of points and impose unitarity to hold in every single one of them,
\small\begin{verbatim}
unitn[s_,nmax_]:=ComplexExpand[Re[#1]]^2+ComplexExpand[Im[#1]]^2<=1&/@Sn[s,nmax];
unitarityn[nmax_,ngrid_]:=And@@Flatten[(unitn[#, nmax]&)/@N[sgrid[ngrid],500]];
\end{verbatim}\normalsize
All that is left to do is to define the components $t_1(s)$, $t_2(s)$, $r_1(s)$, since we will be plotting in this space
\small\begin{verbatim}
t1n[nmax_]:=(Sn[2,nmax][[2]]+Sn[2,nmax][[4]])/2;
t2n[nmax_]:=(Sn[2,nmax][[1]]-Sn[2,nmax][[2]])/(2*n)
            +(Sn[2,nmax][[3]]-Sn[2,nmax][[4]])/(2*n);
r1n[nmax_]:=(Sn[2,nmax][[2]]-Sn[2,nmax][[4]])/2;
\end{verbatim}\normalsize
With everything defined, we are now in position to construct the main function
\small\begin{verbatim}
fsolradialn[theta_,phi_,nmax_,ngrid_]:=fsolradialn[theta,phi,nmax,ngrid]=
FindMaximum[{R,And[unitarityn[nmax,ngrid],{t1n[nmax],t2n[nmax],r1n[nmax]}== 
R*{Sin[theta]Cos[phi],Sin[theta]Sin[phi],Cos[theta]}]},varsn[nmax]~Join~{R},
MaxIterations -> 10000, Method -> "MOSEK"];
\end{verbatim}\normalsize
which returns a point on the monolith's boundary in the direction defined by polar angle $\theta$ and azimuthal angle $\phi$. By choosing different angles we can generate the allowed regions in the space of the S-matrices.

\subsection{Dual}
In the dual approach we start in a similar fashion, by defining the ansatz of the dual variable
\small\begin{verbatim}
v1={0,1/4,0,1/4,1/4,1/4}; v2={1/(4n),-(1/(4n)),1/(4n),-(1/(4n)),1/4,-(1/4)};
v3={1/(4n),(-1+n)/(4n),-(1/(4n)),1/4(-1+1/n),0,0};
Kdn[s_,nmax_]:=Kdn[s,nmax]=2/(Sqrt[s]Sqrt[4-s]*(s-2))*
(a1*v1+a2*v2+a3*v3+Sum[{singP[n],adjP[n],singM[n],adjM[n],sym[n],anti[n]}
*r[2,s]^n+Transpose[c].{singP[n],adjP[n],singM[n],adjM[n],sym[n],anti[n]}
*r[2,4-s]^n,{n,1,nmax}]);
varsdualn[nmax_]:=Variables[Kdn[3.5,nmax]];
\end{verbatim}\normalsize
where $\mathbf{v_i}$ are, as previously mentioned, eigenvalues of $\mathcal{C}^T$ with eigenvalue $1$ and the coefficients $a_i$ determine the direction in the space of the scattering amplitudes where the point will be.

Next, we need to evaluate the integral of (\ref{eq:dual functional definition}), which we will want to minimize. Here we chose to perform the integration by Chebyshev quadrature, which can be implemented in the following way
\small\begin{verbatim}
gridn[nintpts_,precision_]:=Table[N[Cos[j*\[Pi]/(nintpts+1)],precision],
{j,1,nintpts}];
integralsn[nintpts_,precision_]:=integralsn[nintpts,precision]=
2*Table[Expand[Times@@(x-Drop[gridn[nintpts,precision],{k}])/
Times@@(gridn[nintpts,precision][[k]]-Drop[gridn[nintpts,precision],{k}])]
/.x^(m_.):> Boole[EvenQ[m]]/(m+1)//N[#,precision]&,{k,nintpts}];
fn[x_,nmax_]:=-(2/\[Pi])*((16(-1+x))/(1+x)^3)*Total@Abs@Kdn[(8(1+x^2))/
(1+x)^2,nmax];
goaln[nmax_,nintpts_,precision_]:=goaln[nmax,nintpts,precision]=((fn[#,nmax]&)
/@gridn[nintpts,precision]//ExpandAll//Chop).integralsn[nintpts,precision];
\end{verbatim}\normalsize
In the above functions we can choose the number of points \textit{nintpts} with which we numerically evaluate the integral and the \textit{precision} of this procedure, which should be big enough. At the end the \texttt{goal} function provides us with the integral's expression, which must be minimized. 
Before that, however, we define an auxiliary function
\small\begin{verbatim}
rep[\[Theta]_,\[Phi]_]:=First[Solve[Sin[\[Theta]]Cos[\[Phi]]*a1
                       +Sin[\[Theta]]Sin[\[Phi]]*a2+Cos[\[Theta]]*a3==1]];
\end{verbatim}\normalsize
which ensures that we are minimizing along the direction specified by the angles $\theta$, $\phi$.
Finally, we construct the function that minimizes $\mathcal{F}_d [\mathbf{K}(s)]$, namely
\small\begin{verbatim}
solnM[theta_,phi_,nmax_,nintpts_,precision_]:=solnM[theta,phi,nmax,nintpts,
precision]=Module[{angsol=rep[theta, phi],var},var=First[Intersection[
{a1,a2,a3},{angsol[[1,1]]}]];FindMinimum[goaln[nmax,nintpts,precision]
/.angsol,DeleteCases[varsdualn[nmax], var],MaxIterations->1000,
Method->"MOSEK"]//ReplacePart[#,2->Sort[Join[#[[2]],{var->angsol[[1,2]]
/.#[[2]]}]]]&];
\end{verbatim}\normalsize
which returns a point in the space of scattering amplitudes. Repeating this for several angles we obtain a surface, where everything outside of it is excluded, like the ones in figures~\ref{fig:N2monos} and \ref{fig:N7monos}. However, we point out that in order to generate such smooth figures this code takes around an hour, in spite of using \texttt{MOSEK}. A rougher, but still reasonable image with less points, can nonetheless be obtained in a few minutes. Increasing the free parameters \textit{nmax}, \textit{nintpts} and \textit{precision} it is possible to check that the obtained surfaces tend to the optimal bounds and converge to surfaces obtained with the primal approach, thus implying a vanishing duality gap.

\newpage

\bibliographystyle{JHEP}
\bibliography{main}

\end{document}